\documentclass[11pt,letterpaper]{article}
\pdfoutput=1
\usepackage{jheppub}
\usepackage{amsfonts}
\usepackage{amsmath}
\allowdisplaybreaks[4]         
\usepackage{amssymb}   
\usepackage{euscript}       
\usepackage{color}    
\usepackage{comment}   
\usepackage[header,title,page,titletoc]{appendix}
\usepackage{subcaption}

\newtheorem{theorem}{Theorem}

\newtheorem{defi}{Definition}

\usepackage{tensor}
\usepackage{braket}
\usepackage{float}
\newcommand{\be}{\begin{equation}}
\newcommand{\ee}{\end{equation}}

\newcommand{\bea}{\begin{eqnarray}}
\newcommand{\eea}{\end{eqnarray}}
\newcommand{\ba}{\begin{eqnarray}}
\newcommand{\ea}{\end{eqnarray}}

\newcommand{\beq}{\begin{equation}}
\newcommand{\eeq}{\end{equation}}
\newcommand{\beqa}{\begin{eqnarray}}
\newcommand{\eeqa}{\end{eqnarray}}
\newcommand{\beqar}{\begin{eqnarray*}}
\newcommand{\eeqar}{\end{eqnarray*}}

\newcommand{\req}[1]{(\ref{#1})}

\newcommand{\C}{\mathcal{C}}

\newcommand{\diff}{d}

\newcommand{\p}{\partial}

\renewcommand{\href}[2]{#2}

\title{Electromagnetic Quasitopological Gravities}
\author[a]{Pablo A. Cano} 
\author[b]{and \'Angel Murcia}

\affiliation[a]{Instituut voor Theoretische Fysica, KU Leuven.\\
Celestijnenlaan 200D, B-3001 Leuven, Belgium}

\affiliation[b]{Instituto de F\'isica Te\'orica UAM/CSIC.\\C/ Nicol\'as Cabrera, 13-15, C.U. Cantoblanco, 28049 Madrid, Spain\vspace{0.1cm}}

\vspace{0.2cm}
\emailAdd{pabloantonio.cano@kuleuven.be}
\emailAdd{angel.murcia@csic.es} 

\abstract{We identify a set of higher-derivative extensions of Einstein-Maxwell theory that allow for spherically symmetric charged solutions characterized by a single metric function $f(r)=-g_{tt}=1/g_{rr}$. These theories are a non-minimally coupled version of the recently constructed Generalized Quasitopological gravities and they satisfy a number of properties that we establish. We study magnetically-charged black hole solutions in these new theories and we find that for some of them the equations of motion can be fully integrated, enabling us to obtain analytic solutions. In those cases we show that, quite generally, the singularity at the core of the black hole is removed by the higher-derivative corrections and that the solution describes a globally regular geometry. 
In other cases, the equations are reduced to a second order equation for $f(r)$. Nevertheless, for all the theories it is possible to study the thermodynamic properties of charged black holes analytically. We show that the first law of thermodynamics holds exactly and that the Euclidean and Noether-charge methods provide equivalent results. We then study extremal black holes, focusing on the corrections to the extremal charge-to-mass ratio at a non-perturbative level. We observe that in some theories there are no extremal black holes below certain mass. We also show the existence of theories for which extremal black holes do not represent the minimal mass state for a given charge. The implications of these findings for the evaporation process of black holes are discussed.}

\begin{document}
\maketitle
\flushbottom
\newpage

\section{Introduction}
Higher-derivative theories of gravity are modifications of General Relativity inspired by high-energy physics. In fact, it is usually accepted that the effective action of a UV-complete theory of gravity will contain an infinite tower of higher-derivative terms --- in particular this is the case of String Theory \cite{Gross:1986mw,Gross:1986iv,Bergshoeff:1989de}. 
%it is a common prediction of quantum gravity theories, such as string theory, that this type of terms will appear in the low energy effective action of gravity. 
However, in recent years, there has been a genuine interest in studying higher-derivative gravities from a bottom-up approach, regardless of their possible fundamental origin. %This is so for several reasons. 
Among the many reasons for this interest, let us mention a few.

On the one hand, higher-order gravities are known to have an improved UV behaviour with respect to GR \cite{Stelle:1976gc}, and hence it is interesting to explore which features of Einstein's theory are modified when one takes into account these corrections. This is especially appealing in situations of large curvature such as the early universe or black holes, \textit{e.g.} \cite{Starobinsky:1980te,Wheeler:1985nh,Boulware:1985wk,Myers:1988ze,quasi2,quasi,Lu:2015cqa}.
%The modifications to black holes or the early universe can be especially interesting. 
Another fascinating application of higher-derivative gravities is found in the context of the AdS/CFT correspondence \cite{Maldacena,Witten,Gubser}, which allows us to learn about Conformal Field Theories by studying gravity theories in Anti-de Sitter space. In this respect, higher-order gravities can be used to probe a larger set of dual CFTs than GR; a strategy that has sometimes led to outstanding results, \textit{e.g.} \cite{Buchel:2008vz,Myers:2008yi,Camanho:2009vw,deBoer:2009pn,Camanho:2010ru,Myers:2010tj,Myers:2010xs,Bueno1,Bueno2}.
Yet there is also a phenomenological motivation to study higher-derivative theories.
%Yet another motivation to study higher-derivative theories has to do with gravitational waves. 
Due to the increasingly accurate measurements of gravitational waves coming from black-hole and neutron-star mergers \cite{LIGOScientific:2018mvr}, we are at the verge of testing GR with far more precision than ever. Thus, there is a growing interest in searching for deviations from GR, including the presence of higher-derivative corrections in astrophysical back holes --- see \textit{e.g.} \cite{Cardoso:2009pk,Blazquez-Salcedo:2017txk,Berti:2018vdi,Cardoso:2018ptl,Okounkova:2019zjf,Sennett:2019bpc,Carson:2020ter,Cano:2020cao}.

Despite their interesting properties and applications, one important consideration when working with higher-order gravities is the (im)possibility of performing computations. As theories with terribly complicated equations of motion, it is not always the case that one can say anything relevant about them. For instance it is usually not possible to find explicit static spherically symmetric solutions, and in most cases even the numeric resolution is challenging due to the higher-order derivatives in the equations. This makes the study of generic theories a very ungrateful task. On the contrary, there are some special theories that are amenable to computations and these are very appealing at least from a practical point of view. 

One of the most well-known higher-derivative extensions of GR is Lovelock gravity \cite{Lovelock1,Lovelock2,Padmanabhan:2013xyr}. Being the only higher-curvature theories with second-order equations, Lovelock gravities have been thoroughly studied in the literature --- regarding black hole solutions, see \textit{e.g.} Refs. \cite{Wheeler:1985nh,Boulware:1985wk,Cai:1998vy,Cai:2001dz,Camanho:2011rj}. They have an important limitation, though, since they are highly constrained by the spacetime dimension. Thus, the quadratic Lovelock invariant --- corresponding to the Gauss-Bonnet density --- only becomes dynamical for $D\ge 5$, and in general one needs to move to $D\ge 2n+1$ if one wants the $n-$th order Lovelock density to become non-trivial. 
Although it is not possible to circumvent Lovelock's theorem, one possible strategy to find interesting theories consists in asking that they possess second-order equations of motion on certain situations. This idea gave rise to the so called Quasitopological gravity \cite{quasi2,quasi}, as a cubic curvature density such that the traced equations of motion, as well as the equations for static and spherically symmetric metrics, are of second order. It was additionally checked that the linearized equations on constant curvature backgrounds are proportional to the linearized Einstein tensor \cite{quasi}. Higher-order versions of Quasitopological gravity have been later constructed \cite{Dehghani:2011vu,Cisterna:2017umf,Bueno:2019ycr}. However, again these theories are constrained by the dimension of spacetime as they only exist in $D\ge 5$.

More recently, a more general class of theories containing Quasitopological and Lovelock gravities has been identified. The theories of this class are known as Generalized Quasitopological gravities (GQGs) \cite{Hennigar:2017ego,PabloPablo3}, and their main characteristic is that they allow for static and spherically symmetric (SSS) solutions of the form
\begin{equation}
ds^2=-f(r)dt^2+\frac{dr^2}{f(r)}+r^2d\Omega^2_{(D-2)}\, ,
\end{equation}
this is, with $g_{tt}g_{rr}=-1$. From their defining property, it can also be proven that the equation of motion of $f(r)$ is partially integrable and that the linearized equations on constant curvature backgrounds are of second order. In addition, it is verified in all the known examples that black hole thermodynamics can be studied analytically in these theories. There also exist more refined subsets of GQGs that possess single-function Taub-NUT solutions \cite{Bueno:2018uoy} and/or second-order cosmological Friedmann equations \cite{Arciniega:2018fxj,Cisterna:2018tgx,Arciniega:2018tnn}. Unlike Quasitopological gravities, their generalized counterparts do exist in $D=4$, and in fact, they have been shown to exist in all dimensions and at all orders \cite{Bueno:2019ycr}. In addition, they are more general than one would expect, and it turns out that they provide a basis for at least all operators in the sets $\{$Riem$^n\,| n\in \mathbb{N}_0\}$ and  $\{\text{Riem}^n\nabla \text{Riem}\nabla\text{Riem}\,| n\in \mathbb{N}_0 \}$  in the effective action of gravity once field redefinitions are taken into account \cite{Bueno:2019ltp}.

The identification of GQGs was triggered by the earlier construction of Einsteinian cubic gravity \cite{PabloPablo} as a theory with second-order linearized equations on constant curvature backgrounds in arbitrary dimension. The special form of black hole solutions in this theory was soon noticed \cite{Hennigar:2016gkm,PabloPablo2} and this motivated the definition of Generalized Quasitopological gravities established in Refs.~ \cite{Hennigar:2017ego,PabloPablo3}. 
Since then, many applications of these theories have been discussed in the literature in a wide range of topics including cosmology, black holes, holography and phenomenology \cite{Ahmed:2017jod,PabloPablo4,Feng:2017tev,Hennigar:2017umz,Hennigar:2018hza,Poshteh:2018wqy,ECGholo,Bueno:2018yzo,Mehdizadeh:2019qvc,Mir:2019ecg,Mir:2019rik,Erices:2019mkd,Cano:2019ozf,Burger:2019wkq,Frassino:2020zuv,Bueno:2020odt,Quiros:2020uhr,Adair:2020vso,Pookkillath:2020iqq,Khodabakhshi:2020hny,Edelstein:2020nhg,Konoplya:2020jgt}.

The definition of GQGs does not involve matter, so as an important extension one may consider coupling other fields to these theories. This is a nice exercise if the matter fields also respect the property of producing SSS solutions satisfying $g_{tt}g_{rr}=-1$ \cite{Salgado:2003ub}. A simple and relevant example of this is provided by a minimally coupled Maxwell field, a case which has been explored in Refs.~\cite{PabloPablo2,Hennigar:2017umz,Cano:2019ozf,Frassino:2020zuv} in the context of Einsteinian cubic gravity and higher-order GQGs.  It is also a satisfactory strategy to couple these theories to non-linear electrodynamics, as recently shown in \cite{KordZangeneh:2020qeg}. However, these examples only involve minimally coupled gauge fields, which is a very restricted way of coupling a vector to gravity. In general, higher-derivative effective actions may contain all types of couplings between the fields present in them. In addition, non-minimal couplings could have interesting effects that do not show up in the minimally coupled case, so they may be worth exploring. 
On the other hand, the study of charged black holes with higher-derivative corrections is a current matter of research in the context of string phenomenology due to its relation with the Weak Gravity Conjecture (WGC) \cite{ArkaniHamed:2006dz} --- see \textit{e.g.} \cite{Cheung:2018cwt,Hamada:2018dde,Bellazzini:2019xts,Charles:2019qqt,Loges:2019jzs,Goon:2019faz,Cano:2019oma,Cano:2019ycn,Andriolo:2020lul,Loges:2020trf}. In this respect, it may be interesting to analyze the effect of vector-curvature couplings on charged black holes, since those couplings indeed appear in stringy effective actions \cite{Baron:2017dvb,Eloy:2020dko,Elgood:2020xwu,Ortin:2020xdm}.

Thus, one interesting question is whether it is possible to find a family of theories analogous to Generalized Quasitopological gravities in the case of non-minimal couplings between the curvature and a gauge field. The goal of this paper is to show that such generalization is indeed possible and to study charged black hole solutions in the new theories. In particular, we show that a satisfactory extension can be achieved by searching for theories whose magnetically-charged static spherically symmetric solutions satisfy the condition $g_{tt}g_{rr}=-1$. Note that, even though we focus on theories with ``simple'' magnetic solutions, one can generate theories with analogous electric solutions by dualizing the vector field.  We obtain an infinite number of Lagrangians belonging to this new class of theories, that we denote Electromagnetic Generalized Quasitopological gravities (EGQG)\footnote{Not to be confused with the recently constructed ``quasitopological electromagnetism'' of Refs.~\cite{Liu:2019rib,Cisterna:2020rkc}, which provides a non-linear extension of Maxwell's electromagnetism with interesting properties and with explicit black hole solutions when it is minimally coupled to gravity.}. As we show, for all of these theories it is possible to study the thermodynamic properties of black holes analytically, and for some of them we can even write exact black hole solutions. The paper is organized as follows.
\begin{itemize}
\item In Section~\ref{sec:gen} we review some basic aspects of general  $\mathcal{L}(R_{\mu\nu\rho\sigma},F_{\alpha\beta})$ theories, namely, equations of motion, duality transformations, conserved charges and first law of black hole mechanics. 
\item In Section~\ref{sec:SSS} we analyze the equations of motion of particular $\mathcal{L}(R_{\mu\nu\rho\sigma},F_{\alpha\beta})$ theories for charged static spherically symmetric configurations and we establish the definition as well as the properties of Electromagnetic Generalized Quasitopological gravities. 
\item In Section~\ref{sec:EQG} we study the ``quasitopological'' subset of EGQGs, corresponding to those theories for which the equation of motion of the metric function $f(r)\equiv-g_{tt}$ is fully integrable. We find two infinite families of Lagrangians belonging to this class and we exactly solve the equations for magnetically charged spherically symmetric solutions. We show that in many cases the solutions are non-singular, corresponding to regular black holes or smooth horizonless geometries. We then study the thermodynamic properties of black holes in these theories, showing that the first law of black hole mechanics holds exactly. In addition, we analyze the properties of extremal black holes. 
\item In Section~\ref{sec:EGQG} we study an infinite family of proper Electromagnetic Generalized Quasitopological gravities, \textit{i.e.}, those for which the equation for $f(r)$ is not algebraic. We analyze how this equation can be solved to search for black hole solutions and we manage to determine analytically the thermodynamic properties of these black holes. The extremal limit is also discussed.
\item We discuss our findings and point out new directions in Section~\ref{sec:con}.
\end{itemize}
In addition, we include two appendices with some technical results.

\subsubsection*{Note on acronyms}
\begin{itemize}
\item QG $=$ Quasitopological Gravity\footnote{For some people QG $=$ Quantum Gravity.}
\item GQG $=$ Generalized Quasitopological Gravity
\item EQG $=$ Electromagnetic Quasitopological Gravity
\item EGQG $=$ Electromagnetic Generalized Quasitopological Gravity
\end{itemize}
Sometimes we remove the last ``G'' in these acronyms and we write things like ``GQ theories'', which means ``Generalized Quasitopological theories'', etc.

\section{Some generalities on $\mathcal{L}(R_{\mu\nu\rho\sigma},F_{\alpha\beta})$ theories}\label{sec:gen}
As a preliminary step before focusing on the main topic of this paper, in this section we review some aspects of general  $\mathcal{L}(R_{\mu\nu\rho\sigma},F_{\alpha\beta})$ theories that will be useful for later purposes. 

\subsection{Equations of motion}

Let us consider a general gauge- and diffeomorphism-invariant theory for the metric tensor $g_{\mu\nu}$ and a $\mathrm{U}(1)$ gauge field $A_{\mu}$.  The Lagrangian of such theory must be constructed from contractions of the Riemann curvature tensor $R_{\mu\nu\rho\sigma}$ and the field strength\footnote{The most natural way to achieve a gauge-invariant theory is to impose that the whole dependence of the Lagrangian on the gauge field $A$ takes place through its curvature 2-form $F$.} $F=dA$ using the (inverse) metric $g^{\mu\nu}$, and we denote it by $\mathcal{L}(R_{\mu\nu\rho\sigma},F_{\alpha\beta})$.\footnote{For simplicity we assume no  covariant derivatives acting on Riemann curvature tensors nor on field strengths.}  Then, we define the action by 
\begin{equation}
I[g,A]=\frac{1}{16\pi G} \int_M d^n x \sqrt{\vert g\vert}\mathcal{L}(R_{\mu\nu\rho\sigma},F_{\alpha\beta})\, ,
\label{eq:lrftg}
\end{equation}
where we consider an arbitrary spacetime dimension $n$. Since the results obtained in this subsection are valid for any dimension, we will keep $n$ to be arbitrary for now, but we would like to emphasize that the rest of the manuscript will refer to 4-dimensional theories, as it will become apparent. Also, from now on we set $G=1$. 
 
If we consider pure theories of gravity (that is, with no coupling to electromagnetism), a general formula is known for the associated equations of motion --- see \textit{e.g.} \cite{Padmanabhan:2013xyr}. Our first task in this document shall be to derive an analogous formula in the case of $\mathcal{L}(R_{\mu\nu\rho\sigma},F_{\alpha\beta})$ theories. 
In order to obtain the generalized Einstein's equations, we consider the variation of the action \eqref{eq:lrftg} and evaluate it on a vector of the form $(\delta g^{\mu \nu},0)$: 
\begin{equation}
\begin{split}
\delta I[g,A] (\delta g^{\mu \nu},0)&=\frac{1}{16\pi} \int_M d^n x \sqrt{\vert g \vert }\left\lbrace -\frac{1}{2}g_{\mu \nu}\delta g^{\mu \nu}  \mathcal{L} +\frac{\p  \mathcal{L}}{\p g^{\mu \nu}}\delta g^{\mu \nu}+\frac{ \p \mathcal{L}}{\p R_{\alpha \beta \rho \gamma}}  \delta  R_{\alpha \beta \rho \gamma} \right\rbrace,
\end{split}
\label{eq:vmga}
\end{equation}
where the term $\delta  R_{\alpha \beta \rho \gamma}$ is evidently not independent from $\delta g^{\mu \nu}$. Defining, for convenience of notation,
\begin{equation}
P^{\alpha \beta \rho \gamma}=\frac{ \p \mathcal{L}}{\p R_{\alpha \beta \rho \gamma}}\, ,\quad \mathcal{M}^{\alpha \beta}=-\frac{1}{2}\frac{\p \mathcal{L}}{\p F_{\alpha \beta}}\, ,
\label{eq:defpandm}
\end{equation}
it is possible to find that, up to total derivatives, 
\begin{equation}
P^{\alpha \beta \rho \gamma}\delta R_{\alpha \beta \rho \gamma}=-2 \nabla^\sigma \nabla^\beta P_{\mu \sigma \beta \nu} \delta g^{\mu \nu }-\tensor{P}{_{\beta}^{\sigma \mu \nu}} R_{\rho \sigma \mu \nu} \delta g^{\rho \beta}\,. 
\end{equation}
On the other hand, let us work out in two different manners the Lie derivative $L_\xi \mathcal{L}$  with respect to an arbitrary vector field $\xi \in \mathfrak{X}(M)$:
\begin{eqnarray}
L_\xi \mathcal{L}&=&\xi^\mu \nabla_\mu \mathcal{L}=\xi^\mu P^{\nu \rho \sigma \beta} \nabla_\mu R_{\nu \rho \sigma \beta}-2\xi^\mu \mathcal{M}^{\alpha \beta} \nabla_\mu F_{\alpha \beta} \, ,\\
L_\xi \mathcal{L}&=&P^{\nu \rho \sigma \beta} L_\xi R_{\nu \rho \sigma \beta} +\frac{\p \mathcal{L}}{\p g_{\alpha \beta}}L_\xi g_{\alpha \beta}-2\mathcal{M}^{\alpha \beta} L_\xi F_{\alpha \beta}\,.
\end{eqnarray}
If we take into account that
\begin{eqnarray}
P^{\nu \rho \sigma \beta }L_\xi R_{\nu \rho \sigma \beta}&=&\xi^\mu P^{\nu \rho \sigma \beta} \nabla_\mu R_{\nu \rho \sigma \beta}+4(\nabla_{\mu} \xi_\nu) P^{\mu \rho \sigma \beta} \tensor{R}{^{\nu}_{\rho \sigma \beta}} \, , 
\\ L_\xi g_{\alpha \beta}&=&2\nabla_{(\alpha} \xi_{\beta)} \, , \\
 \mathcal{M}^{\alpha \beta} L_\xi F_{\alpha \beta}&=&\xi^\mu \nabla_\mu F_{\alpha \beta} \mathcal{M}^{\alpha \beta}+2 \nabla_\alpha \xi^\mu F_{\mu \beta} \mathcal{M}^{\alpha \beta}\, ,
\end{eqnarray}
then we find that\footnote{We recall that
\begin{equation*}
\frac{\partial \mathcal{L}}{\partial g^{\mu \nu}}=-g_{\mu \alpha}\frac{\partial \mathcal{L}}{\partial g_{\alpha \beta}} g_{\nu \beta}\,.
\end{equation*}}
\begin{equation}
\frac{\p \mathcal{L}}{\p g^{\mu \nu}}=2\tensor{P}{_\mu^{\alpha \beta \gamma}} R_{\nu \alpha \beta \gamma}-2\tensor{\mathcal{M}}{_\mu^\alpha} F_{\nu \alpha}\,.
\label{eq:plpg}
\end{equation}
Consequently, equation \eqref{eq:vmga} may be re-expressed as
\begin{align}
\notag
\delta I[g,A] (\delta g^{\mu \nu},0)&=\frac{1}{16\pi}\int_M d^n x \sqrt{\vert g \vert }\left\lbrace -\frac{1}{2}g_{\mu \nu}\delta g^{\mu \nu}  \mathcal{L} + 2\tensor{P}{_\mu^{\alpha \beta \gamma}} R_{\nu \alpha \beta \gamma}\delta g^{\mu \nu} -2 \tensor{\mathcal{M}}{_\mu^\alpha} F_{\nu \alpha}\delta g^{\mu \nu}\right.\\&  \left. -2 \nabla^\sigma \nabla^\beta P_{\mu \sigma \beta \nu} \delta g^{\mu \nu }-\tensor{P}{_{\mu}^{\sigma \alpha \beta }} R_{\nu \sigma \alpha \beta } \delta g^{\mu \nu}\right\rbrace,
\end{align}
Hence the gravitational equations of motion of any theory given by the action \eqref{eq:lrftg}, representing the most general theory of gravity  coupled to electromagnetism which includes all possible terms constructed out of curvature tensors, metrics and field strengths, are given by the formula\footnote{In the case of pure gravity we do not need to include an explicit symmetrization in the $\mu\nu$ indices because the terms $\tensor{P}{_{\mu}^{\rho \sigma \gamma}} \tensor{R}{_{\nu \rho \sigma \gamma}}$ and $ \nabla^\sigma \nabla^\rho P_{\mu \sigma\nu\rho}$ are symmetric \cite{Padmanabhan:2013xyr}. However, in the case at hands we were not able to prove that the different structures appearing in $\mathcal{E}_{\mu\nu}$ are automatically symmetric --- although we highly suspect it --- and hence we have to symmetrize explicitly.}
\begin{equation}\label{eq:EinsteinEq}
\mathcal{E}_{\mu\nu}=\tensor{P}{_{(\mu}^{\rho \sigma \gamma}} \tensor{R}{_{\nu) \rho \sigma \gamma}} -\frac{1}{2}g_{\mu \nu} \mathcal{L}+2 \nabla^\sigma \nabla^\rho P_{(\mu| \sigma|\nu)\rho}-2\tensor{\mathcal{M}}{_{(\mu}^\alpha} \tensor{F}{_{\nu) \alpha}}=0\,.
\end{equation}
On the other hand, the derivation of the generalized Maxwell's equation is straightforward and it yields
\begin{equation}\label{eq:MaxwellEq}
\mathcal{E}^{\nu}=\nabla_{\mu}\mathcal{M}^{\mu\nu}=0\,.
\end{equation}

\subsection{Duality transformations}\label{sec:dual}
A duality transformation of a $\mathrm{U}(1)$ gauge field in four dimensions has the effect of exchanging the initial vector field by a dual vector field whose equations of motion correspond to the Bianchi identity of the former, and viceversa.  Intuitively, this operation corresponds to the exchange of electric and magnetic fields.  An interesting property of (Einstein)-Maxwell's theory --- and also of many extensions of it, such as $\mathcal{N}=2$, $d=4$ SUGRA --- is the fact that it is invariant under duality transformations. However, in general the dual theory does not need to coincide with the original one, and thus the duality transformation establishes a map between two different theories.\footnote{Let us note that it is possible to define more general notions of electromagnetic duality than the one we are considering here.  A generic duality transformation establishes an isomorphism between two different theories, that is, a bijection between their configuration and solution spaces \cite{Hull:1995gk,Lazaroiu:2016iav,Lazaroiu:2020vne}, but a remarkable property of certain theories --- such as ungauged four-dimensional supergravity --- is the fact that they admit a non-trivial subgroup (the U-duality group) of the duality group that does preserve the theory \cite{Gaillard:1981rj}. We thank Carlos S. Shahbazi for clarifying these points to us.} This is usually the case when one considers non-minimal couplings between the curvature and the field strength.

 Let us show how the transformation works on a general $\mathcal{L}(R_{\mu\nu\rho\sigma},F_{\alpha\beta})$ theory. We consider the action
\begin{equation}\label{eq:dualth}
I=\frac{1}{16\pi}\int d^4x\sqrt{|g|}\mathcal{L}(R_{\mu\nu\rho\sigma},F_{\alpha\beta})\, ,\quad\text{where}\quad F_{\mu\nu}=2\partial_{[\mu}A_{\nu]}\, .
\end{equation}
Then, we have to introduce an auxiliary field $B_{\mu}$ whose equation of motion yields the Bianchi identity $dF=0$, which is locally equivalent to $F=dA$. Thus, we add the term $B_{\mu}\nabla_{\nu}F_{\alpha\beta}\epsilon^{\mu\nu\alpha\beta}$ in the action and now the variables are $F$ (instead of $A$) and $B$. Integrating that term by parts we can write
\begin{equation}
I_{\mathrm{dual}}=\frac{1}{16\pi}\int d^4x\sqrt{|g|}\left[\mathcal{L}(R_{\mu\nu\rho\sigma},F_{\alpha\beta})-2F^{\mu\nu}\star G_{\mu\nu}\right]\, ,
\label{eq:dualacth}
\end{equation}
where $G_{\mu\nu}=2\partial_{[\mu}B_{\nu]}$ and 
\begin{equation}
\star G_{\mu\nu}=\frac{1}{2}\epsilon_{\mu\nu\alpha\beta}G^{\alpha\beta}
\end{equation}
is the dual field strength. Now, as required, the variation with respect to $B$ yields the Bianchi identity of $F$,  and the variation with respect to $F$ gives a relation between $F$ and $G$, 
\begin{equation}\label{eq:dualfield}
G_{\mu\nu}=-\frac{1}{2}\star \frac{\partial \mathcal{L}}{\partial F^{\mu\nu}}\, .
\end{equation}
This is a direct expression of the dual vector field $G$ , but now we have to invert this expression to obtain $F$, which in general is some complicated function of $G$ and of the curvature, 
\begin{equation}
F_{\mu\nu}=F_{\mu\nu}\left(\star G_{\alpha\beta}, R_{\rho\sigma\lambda\gamma}\right)\, .
\end{equation}
Inserting this back in \req{eq:dualacth} we get the dual theory, which on general grounds is different to the original theory. 
In appendix \ref{app:dualEGQG} we compute explicitly the dual theory in the case of a Lagrangian quadratic in $F$. 

\subsection{Mass, charges and thermodynamics}
For the purposes of this paper, let us very briefly review the issue of conserved charges and black hole thermodynamics. In the case of electric and magnetic charges, these are obtained directly from the equations of motion. In particular, the generalized Maxwell equation can be written as
\begin{equation}
d\star \mathcal{M}=0\, ,\quad \text{where}\quad \mathcal{M}=-\frac{1}{4}\frac{\partial \mathcal{L}}{\partial F^{\mu\nu}}dx^{\mu}\wedge dx^{\nu}\, .
\end{equation}
If a current three-form $J$ is placed on the right-hand-side, the equation implies that the current is conserved, $dJ=0$, and hence the natural definition of electric charge is 

\begin{equation}
Q=\frac{1}{4\pi}\int_{\mathbb{S}^2_{\infty}} \star \mathcal{M}\, ,
\end{equation}
where the integration is taken at spatial infinity. Let us note that in asymptotically flat spacetimes, with a Lagrangian $\mathcal{L}=R-F^2+$higher-order, we have $\mathcal{M}\rightarrow F$ asymptotically, so in practice we can replace $\mathcal{M}$ by $F$ as long as the integral is performed at infinity.  In the asymptotically AdS case, there is a theory-dependent constant $c_q$ such that $\mathcal{M}\rightarrow c_q F$ at the boundary of AdS.  On the other hand, the magnetic charge is defined in the standard way,

\begin{equation}
P=\frac{1}{4\pi}\int_{\mathbb{S}^2_{\infty}} F\, .
\end{equation}

Gravitational conserved charges in higher-order gravities were studied in Refs.~\cite{Deser:2002jk,Senturk:2012yi,Adami:2017phg}, but here we are only interested in the total mass. In the case of asymptotically flat spacetimes, it turns out that the mass can be formally computed using the same prescriptions as for GR, for instance via the ADM \cite{arnowitt:1961zz,Arnowitt:1962hi} or the Abbott-Deser \cite{Abbott:1981ff} formulas.  Thus, the mass can be computed by examining the asymptotic behaviour of the metric in the usual way, \textit{i.e.}, identifying the term $2GM/r\in g_{tt}$. In the AdS case, a global theory-dependent factor should be added to those formulas, corresponding to replacement of Newton's constant by the so-called effective Newton's constant $G_{\rm eff}$ \cite{Aspects}. 
For $\mathcal{L}(R_{\mu\nu\rho\sigma},F_{\alpha\beta})$ theories we do not expect these results to be affected, since the higher-order operators formed from $F_{\alpha\beta}$ decay too fast at infinity to contribute to the mass. 

Regarding black hole thermodynamics, it is known that higher-curvature gravities minimally coupled to a Maxwell Lagrangian $F^2$ satisfy the first law of black hole mechanics \cite{Bardeen:1973gs,Gao:2003ys}
\begin{equation}\label{eq:1stlaw1}
dM=T dS+\Phi_h dQ+\Psi_h dP\, .
\end{equation}
Here $M$, $Q$ and $P$ are the mass and charges computed as we specified above, $T$ is the Hawking temperature of the black hole \cite{Hawking:1974sw} and $S$ is Wald's entropy \cite{Wald:1993nt,Iyer:1994ys}, given by\footnote{There are some subtleties when defining the Noether charge for theories that involve fields with internal gauge freedom --- see \textit{e.g.} Refs.~\cite{Prabhu:2015vua,Elgood:2020svt} for recent discussions on this topic.} 
\begin{equation}
S=-2\pi \int_{\Sigma} d^2x\sqrt{h}\frac{\partial \mathcal{L}}{\partial R_{ \mu \nu\rho\sigma}} \epsilon_{\mu \nu}\epsilon_{\rho \sigma}\, ,
\end{equation}
where the integral is carried out on the bifurcation surface of the event horizon and $ \epsilon_{\mu \nu}$ is the binormal to this surface. 
In addition, $\Phi_h$ is the electrostatic potential at the horizon, while $\Psi_h$ is the electrostatic potential of the dual vector field whose field strength is given by \req{eq:dualfield}.  These can be computed according to the following relations
\begin{equation}
\xi^{\nu}F_{\mu\nu}=\partial_{\mu}\Phi\, ,\quad \xi^{\nu}G_{\mu\nu}=\partial_{\mu}\Psi\, ,
\end{equation}
where $\xi^{\nu}$ is the Killing vector that generates the horizon, with the condition that $\Phi$ and $\Psi$ vanish asymptotically.

It was shown in Ref.~\cite{Rasheed:1997ns} that the form of the first law is unchanged when instead of the Maxwell Lagrangian one considers non-linear electrodynamics minimally coupled  to Einstein gravity. However, such analysis does not include the case of non-minimally coupled terms, and one may wonder if the form of the first  law could be altered in that case. This is, just like the entropy is no longer proportional to the area when there are higher-curvature terms, $S\neq A/4$, the question is whether the quantities $\Phi$ or $Q$ in \req{eq:1stlaw1} could have to be replaced by different ones in order for the first law to hold. Another non-trivial question is whether the Noether's charge and the Euclidean path integral approaches yield equivalent results for black hole thermodynamics \cite{Iyer:1995kg,Feng:2015sbw}.

\section{Static and spherically symmetric solutions}\label{sec:SSS}
In this section we address the problem of finding static, spherically symmetric (SSS) solutions of $\mathcal{L}(R_{\mu\nu\rho\sigma},F_{\alpha\beta})$ theories. Obviously, the equations of motion are far too general to be solved without specifying a Lagrangian, so instead our aim is to understand the structure of those equations and to describe a class of theories for which the problem can be simplified.

For a SSS configuration, we can write a general ansatz for the metric in the usual form
\begin{equation}
\diff s_{N,f}^2=-N^2(r) f(r) \diff t^2+\frac{\diff r^2}{f(r)}+r^2 \left(\diff \theta^2+\sin^2\theta\diff \phi^2\right)\, ,
\label{eq:sss1}
\end{equation}
which depends on two functions $N$ and $f$, while in the case of the vector field we will assume either an electric or magnetic ansatz, as given by
\begin{align}
A^{\rm e}&=\Phi(r)dt\quad \Rightarrow \quad F^{e}=-\Phi'(r) dt\wedge dr\, ,\\
A^{\rm m}&=\chi(\theta)d\phi\quad \Rightarrow \quad F^{m}=\chi'(\theta) d\theta\wedge d\phi\, .
\end{align}
One may also consider dyonic vectors, but this increases the difficulty of the problem taking into account the non-linearity of Maxwell equations and that duality invariance is generically lost, so we will discuss purely electric or purely magnetic configurations only. 

\subsection{The reduced Lagrangian}
The easiest way to find the equations of motion consists in evaluating the Lagrangian on this configuration so that we obtain a reduced Lagrangian, defined as
\begin{equation}
L_{N,f,\Phi}=\sqrt{|g|}\mathcal{L}\Big|_{ds^2_{N,f}, A^{e}}\, ,\quad\text{and} \quad L_{N,f,\chi}=\sqrt{|g|}\mathcal{L}\Big|_{ds^2_{N,f}, A^{m}}\, ,
\end{equation}
for the electric and magnetic configurations respectively.  Then, the equations of motion for the variables $N$, $f$ and $\Phi$ in the electric case are given by
\begin{align}
\mathcal{E}_{N}=&\frac{\delta L_{N,f,\Phi}}{\delta N}\, ,\quad
\mathcal{E}_{f}=\frac{\delta L_{N,f,\Phi}}{\delta f}\, ,\quad
\mathcal{E}_{\Phi}=\frac{\delta L_{N,f,\Phi}}{\delta \Phi}\, ,
\end{align}
and similarly in the magnetic case, where $\delta/\delta N$, etc, represent the Euler-Lagrange variation. Using the chain rule, it can be shown that the equations $\mathcal{E}_{N}=\mathcal{E}_{f} =\mathcal{E}_{\Phi}=0$ are equivalent to some components of the equations of motion obtained from direct evaluation of \req{eq:EinsteinEq} and \req{eq:MaxwellEq}. The rest of the equations are then satisfied on account of the Bianchi identities \cite{PabloPablo3}.  An explicit proof of these statements is included in Appendix \ref{app:redaction}.
So far we have made no assumptions on the form of the Lagrangian $\mathcal{L}(R_{\mu\nu\rho\sigma},F_{\alpha\beta})$, besides it being an invariant formed from the curvature, the metric, and the field strength of the vector field. In order to make further progress, we are going to assume that the Lagrangian is a polynomial in $F_{\alpha\beta}$ and $R_{\mu\nu\rho\sigma}$, \textit{i.e}, it is composed of terms of the form $F^{2m}R^n$, with the indices contracted appropriately. 

The discussion proceeds differently for electric or magnetic configurations, so let us study both cases separately.

\subsubsection*{Electric solutions}
Let us first consider the case of an electric vector field. By looking at the structure of the curvature tensor on a SSS metric \cite{Deser:2005pc}, it is not difficult to show that a monomial composed out of $2m$ field strengths and $n$ curvatures has the following structure when evaluated on such configuration,
\begin{equation}
F^{2m}R^n\sim \left(\frac{\Phi'}{N}\right)^{2m}\mathcal{F}(f,f',f'',N,N',N'',r)\, .
\end{equation}
In addition, $\mathcal{F}$ has the property of being homogeneous of degree 0 in $N$, and hence it can be expressed as
\begin{equation}
\mathcal{F}=\sum_{i=0}^{i_{\rm max}}\sum_{j=0}^{j_{\rm max}} \frac{(N')^{i}(N'')^{j}}{N^{i+j}}\mathcal{F}_{ij}(f,f',f'',r)\, ,
\end{equation}
where the sum is always finite, and the functions $\mathcal{F}_{ij}$ are polynomial in $f$, $f'$ and $f''$.  Now, schematically the reduced Lagrangian is $L_{N,f,\Phi}=Nr^2\sin\theta\sum_{n,m}F^{2m}R^n$. Therefore, the Maxwell equation, obtained from variation with respect to $\Phi$, reads
\begin{equation}
\mathcal{E}_{\Phi}=-\frac{d}{dr}\frac{\partial L_{N,f,\Phi}}{\partial \Phi'}=0\, \Rightarrow \frac{\partial L_{N,f,\Phi}}{\partial \Phi'}=Q\, ,
\end{equation}
where $Q$ is an integration constant. Since the left-hand-side of the last equation is a (polynomial) function of $\Phi'$, we can in principle invert it so that we get\footnote{Of course, there may be more than one solution.}
\begin{equation}
\Phi'=\Phi'_{\rm sol}(f,f',f'',N,N',N'',r,Q).
\end{equation}
 In this way we have eliminated one of the variables in the system of equations. Now we have to plug the value of $\Phi'$ in the equations for $N$ and $f$ and we get two differential equations for these functions,
 \begin{equation}
 \mathcal{E}_{N}\Big|_{\Phi'=\Phi'_{\rm sol}}=0\, ,\quad   \mathcal{E}_{f}\Big|_{\Phi'=\Phi'_{\rm sol}}=0\, .
 \end{equation}
 Since $\Phi'_{\rm sol}$ is generically a highly nonlinear (not even polynomial) function of the variables $f$ and $N$ and their derivatives, solving these equations is in general an inaccessible problem. 
 
\subsubsection*{Magnetic solutions}
Let us turn now to the case of magnetic configurations. It can be seen that the monomials $F^{2m}R^n$ have the form 
\begin{equation}
F^{2m}R^n\sim \left(\frac{\chi'(\theta)}{\sin\theta}\right)^{2m}\mathcal{F}(f,f',f'',N,N',N'',r)\, ,
\end{equation}
where $\mathcal{F}$ has the same structure as in the electric case. Then, the reduced Lagrangian has the following form,
\begin{equation}
L_{N,f,\chi}=Nr^2\sin\theta\sum_{n,m}F^{2m}R^n=Nr^2\sin\theta \sum_{n,m}\left(\frac{\chi'(\theta)}{\sin\theta}\right)^{2m}\mathcal{F}^{(n,m)}(f,f',f'',N,N',N'',r)\, ,
\end{equation}
and the Euler-Lagrange equation for $\chi$ reads

\begin{equation}
\mathcal{E}_{\chi}=-\frac{d}{d\theta}\frac{\partial L_{N,f,\chi}}{\partial\chi'}=Nr^2 \sum_{n,m}2m\frac{d}{d\theta}\left(\frac{\chi'(\theta)}{\sin\theta}\right)^{2m-1}\mathcal{F}^{(n,m)}(f,f',f'',N,N',N'',r)=0\, .
\end{equation}
We see that this equation is always solved for $\chi'(\theta)\propto \sin\theta$, and hence we have
\begin{equation}
\chi'(\theta)= P \sin\theta\, ,
\end{equation}
where $P$ is the magnetic charge. Therefore, magnetic vectors are not affected by the corrections and they have the usual form.  Now we can just plug this value of $\chi$ back in the equations of $N$ and $f$. However, let us note that, since $\chi$ does not depend on these functions, in this case we can insert the on-shell value of $\chi$ on the reduced Lagrangian, from where we get a Lagrangian for $N$ and $f$ only. Thus, for magnetic configurations we can consider from the beginning 
\begin{equation}\label{Fmagnetic}
F^m=P d\theta\sin\theta\wedge d\phi\, ,
\end{equation}
and we can derive the equations for $N$ and $f$ from the effective Lagrangian
\begin{equation}
L_{N,f}=Nr^2\mathcal{L}\Big|_{ds^2_{N,f}, F^m}\, ,
\end{equation}
where we are dropping the factor of $\sin\theta$ since it is irrelevant.
In this case, the equations are much simpler than for the electric vector field. 

\subsection{The condition $g_{tt}g_{rr}=-1$: Generalized Quasitopological theories}
Following the previous discussion, we have come to the conclusion that magnetic solutions are much simpler to study than electric ones. However, even if we consider restricting ourselves to these magnetic solutions, the equations for $N$ and $f$ are typically too complicated to obtain relevant information in general, so further simplification is desirable. 

In the case of pure gravity, an interesting class of theories has been identified in recent years. These theories are known as \emph{Generalized Quasitopological gravities}\footnote{The works \cite{Hennigar:2017ego,PabloPablo3} can be considered as the ones establishing the general properties of the new family of theories, but these are motivated by earlier works on Einsteinian cubic gravity \cite{PabloPablo,Hennigar:2016gkm,PabloPablo2}.} (GQGs) \cite{Hennigar:2017ego,PabloPablo3} and they are characterized by possessing SSS solutions of the form 
\begin{equation}
\diff s_{f}^2=- f(r) \diff t^2+\frac{\diff r^2}{f(r)}+r^2 \diff \Omega^2_{(2)}\,,
\label{eq:dsf}
\end{equation}
\textit{i.e.}, with $N=1$, and where in addition the equation of motion for $f$ can be partially integrated. Other remarkable aspects of these theories are that the thermodynamic properties of black holes can be studied fully analytically and that they only propagate a massless graviton on constant curvature backgrounds. 

Interestingly, GQGs can be nicely combined with minimally coupled vector fields, since they respect the property of having solutions with $g_{tt}g_{rr}=-1$. However, one may wonder if it is possible to generalize these theories in the case of non-minimally coupled vector fields. The defining property of GQGs, from where all the rest can be derived\footnote{See also chapter 3 of \cite{CanoMolina-Ninirola:2019uzm} for a refinement on some of the results in \cite{PabloPablo3}.} \cite{Hennigar:2017ego,PabloPablo3}, is that their reduced Lagrangian becomes a total derivative when evaluated on the single-function ansatz \req{eq:dsf}, so we may try to extend this definition when a non-minimally coupled vector is present. The main problem in that case is that the reduced Lagrangian $L_{N,f,A}$ will not in general enjoy the same structure as for pure gravities, since it will strongly depend on the electromagnetic potential $A_{\mu}$. Thus, for instance, it does not seem possible to impose that $L_{1,f,A}$ be a total derivative without specifying the value of $A_{\mu}$. A more reasonable property to ask would be that the Euler-Lagrange equation of $f$ vanishes identically when it is evaluated on $N=1$ and on a gauge field that solves the Maxwell equation,
\begin{equation}\label{eq:EQGelectric}
\left. \frac{\delta L_{1,f,A}}{\delta f}\right \vert_{A=A_{\mathrm{sol}}}=0\,.
\end{equation}
Still, in the electric case we have seen that $A_{\mathrm{sol}}$ typically has a non-polynomial dependence on $f$ and its derivatives, so it seems difficult to find by brute force theories satisfying this property. 

Fortunately, the situation is different for magnetic configurations.  In fact, we have seen that in the magnetic case we can work with a reduced Lagrangian that depends only on $N$ and $f$, since we can set $F=F^{m}$ as in \req{Fmagnetic} from the start. Then, we can see that the reduced Lagrangian $L_{N,f}$ for non-minimally coupled theories $F^{2m}R^n$ with a magnetic vector field has the same structure as for pure gravity theories, and therefore one can extend straightforwardly the definition of Generalized Quasitopological gravities to these non-minimally coupled terms. In particular, we can present the following

\begin{theorem}\label{theo}
Let us consider a theory with a Lagrangian $\mathcal{L}(R_{\mu\nu\rho\sigma},F_{\alpha\beta})$ of the form
\begin{equation}\label{eq:Ltheorem}
\mathcal{L}(R_{\mu\nu\rho\sigma},F_{\alpha\beta})=R-F^2+\text{\emph{higher-derivative terms}},
\end{equation}
where the higher-derivative terms are formed from monomials of the Riemann tensor and the field strength,\footnote{With this we mean that we do not allow terms such as \textit{e.g.} $F^2/R$.}  schematically $R^nF^{2m}$.
 Let us consider a SSS configuration given by the metric \req{eq:dsf} and by a magnetic vector field with field strength \req{Fmagnetic} and let us define the reduced Lagrangian of the system as 
\begin{equation}
L_{f}=r^2\mathcal{L}\Big|_{ds^2_{f}, F^m}.
\end{equation} 
If the  Euler-Lagrange equation for the reduced Lagrangian $L_f$ vanishes identically, \textit{i.e.}, 
\begin{equation}\label{EGQGcond}
\frac{\partial L_f}{\partial f}-\frac{\diff}{\diff r} \frac{\partial L_f}{\partial f' }+\frac{\diff}{\diff r^2} \frac{\partial L_f}{\partial f''}=0\,,
\end{equation}
then the following properties hold:
\begin{enumerate}
\item the theory allows for magnetically-charged SSS solutions of the form \req{eq:dsf}, \req{Fmagnetic},
\item the equation for the function $f$ can be integrated once yielding at most a second-order equation where the mass appears as an integration constant,
\item the only gravitational mode propagated on maximally symmetric backgrounds is the spin-2 massless graviton, and
\item \emph{(Conjecture)} the thermodynamic properties of magnetically-charged static black holes can be obtained analytically.
\end{enumerate}
\end{theorem}

The points 1 and 2 follow immediately from the results in \cite{PabloPablo3} taking into account that the reduced Lagrangian has the same structure as in the case of pure gravity. Point 3 also follows from the results there --- see also \cite{CanoMolina-Ninirola:2019uzm} ---, but it is somewhat trivial, since the terms of the form  $R^nF^{2m}$ with $m>0$ do not contribute to the linearized equations of the metric, while the pure curvature terms satisfying \req{EGQGcond} are known to yield Einstein-like linearized equations. In addition, let us note that the degrees of freedom associated to the gauge field are the same as in Maxwell theory, since the generalized Maxwell equation for any theory of the form \req{eq:Ltheorem} is of second-order in $A_{\mu}$. 
As for point 4, it technically stands as a conjecture since no formal proof has been offered so far. However, we highly suspect a proof must exist due to the large evidence collected in the case of Generalized Quasitopological gravities \cite{Hennigar:2016gkm,PabloPablo2,Hennigar:2017ego,Ahmed:2017jod,PabloPablo4}, and we also show in the next sections that it holds for all the non-minimally coupled theories that we construct.

So far, these results involve magnetically-charged black hole solutions. However, following the procedure explained in Section~\ref{sec:dual}, one can dualize any theory satisfying \req{EGQGcond} and obtain a new theory with electrically-charged solutions of the form \req{eq:dsf}. Hence all the items in the Theorem \ref{theo} hold for the dual theories after replacing ``magnetically-charged'' by ``electrically-charged''.  This motivates the following definition of \emph{Electromagnetic Generalized Quasitopological Gravities} (EGQGs):

\begin{defi}\label{def:EGQG}
We say that a theory $\mathcal{L}(R_{\mu\nu\rho\sigma},F_{\alpha\beta})$ belongs to the family of \emph{Electromagnetic Generalized Quasitopological Gravities (EGQG)} if and only if its Lagrangian or the Lagrangian of its dual theory satisfies the condition \req{EGQGcond}.
\end{defi}

One could refer to these theories as ``Magnetic'' and ``Electric'' GQGs, respectively, but we call them collectively Electromagnetic GQGs for two reasons.  Firstly, because it makes sense to use the adjective \emph{electromagnetic} to express that these theories are non-minimally coupled to an electromagnetic field. Secondly, because theories of one and another class are simply related by duality transformations, and hence they are equivalent. 
Let us also note that, even though the condition \req{EGQGcond} can be satisfied by simple polynomial Lagrangians (see next sections), the dual (electric) theory is typically much more involved and will contain an infinite number of terms, justifying thus why Definition \ref{def:EGQG} was made in terms of the magnetic ansatz \eqref{Fmagnetic}.

EGQGs, just like their pure gravity counterparts\footnote{There is an important qualitative difference between EGQGs and purely gravitational GQGs though. While Lovelock gravities belong naturally to the GQG family, EGQGs do not include Lovelock-like theories in which a gauge field is non-minimally coupled to gravity, like the ones defined at Refs. \cite{Horndeski:1976gi,Yoshida_1991,Feng:2015sbw}.}, come in two different classes. In the general case, the equations of motion of these theories for charged SSS configurations are reduced to a second-order equation for the function $f$. However, there is a special subset of these theories for which the order of this equation is reduced twice again, and we are left with an algebraic equation. In this case we say that the theories belong to the Quasitopological class (without the ``generalized'' part). 
For pure metric theories, Quasitopological gravities only exist $D\ge 5$, but as we show below, infinitely many Electromagnetic Quasitopological gravities exist in $D=4$.

\section{Electromagnetic Quasitopological gravities}\label{sec:EQG}
As we stated above, the family of theories allowing for single-function SSS solutions come in two classes: those for which the equation for $f$ is algebraic are called ``Quasitopological", while those for which $f$ satisfies a 2nd order equation are called ``Generalized Quasitopological".  In this section we focus on the former case. In order to find theories of the EGQG class, one first writes down a general Lagrangian (including for instance all the densities of the form $F^{2m} R^n$ up to a given order), then evaluates the Lagrangian on the configuration given by \req{eq:dsf}, \req{Fmagnetic}, and finally demands that the condition \req{EGQGcond} is satisfied. At the end, that condition gives us a number of constraints on the couplings of the higher-derivative terms. In addition, if we want to restrict to Quasitopological theories, we must only keep the subset of those theories that yield an algebraic equation for $f$. 

At lower orders in the derivative expansion, one can easily find all the theories of this type, but the process becomes more and more involved at higher orders, as the number of independent densities one can include grows very fast. Thus, a general analysis does not seem a priori accessible. Nevertheless, from the analysis of the lower-order densities we can probably extract a general conclusion on the structure of EQGs. Indeed, in the 
appendix \ref{app:a}, we observe the following two facts. First, that there are only two Lagrangians of the form $F^2R$ belonging to the Electromagnetic Quasitopological class\footnote{In this very particular case it even happens that the EGQ family coincides with the quasitopological one. This is of course not a general feature.}. Second, that at order $F^2R^2$, despite the larger number of linearly independent invariants one can construct, there are only two different ways in which these densities modify the equation of $f$. Thus, if we are only interested in studying SSS solutions, it suffices to keep two representative EQ densities at a given order. By repeating an analogous analysis at higher orders, we observe that the situation seems to be general, in the sense that there are many independent EQGs but their equations on SSS metric are degenerate so that there are only two different contributions at every order.  So, instead of studying the whole set of EQGs, we will provide a set of representative theories which --- we conjecture --- span all the possible modifications to SSS solutions. This will be enough for our goal, which is to study black holes in these theories. 

It turns out that it is not difficult to provide a set of two representative Lagrangians of the EQ type at every order.  In order to write them, let us introduce the following notation:
\begin{equation}
\tensor{\left(R^n\right)}{^{\mu\nu}_{\rho\sigma}}=\tensor{R}{^{\mu\nu}_{\alpha_{1}\beta_{1}}}\tensor{R}{^{\alpha_1\beta_1}_{\alpha_{2}\beta_{2}}}\ldots \tensor{R}{^{\alpha_{n-1}\beta_{n-1}}_{\rho\sigma}}\, ,
\end{equation}
with the convention that $\tensor{\left(R^0\right)}{^{\mu\nu}_{\rho\sigma}}=\tensor{\delta}{^{\mu\nu}_{\rho\sigma}}\equiv \tensor{\delta}{^{[\mu}_{[\rho}}\tensor{\delta}{^{\nu]}_{\sigma]}}$.
Then, we have the following Lagrangians of order $R^nF^{2m}$:
\begin{align}
\label{eq:lagegqa}
\mathcal{L}^{(a)}_{n,m}&=\left(2n \tensor{R}{_{\mu}^{\alpha}}\tensor{\delta}{_{\nu}^{\beta}}-(3n-3+4m)\tensor{R}{^{\alpha\beta}_{\mu\nu}}\right)\tensor{\left(R^{n-1}\right)}{^{\mu\nu}_{\rho\sigma}}F^{\rho\sigma}F_{\alpha\beta}\left(F^2\right)^{m-1}\, ,\\ \nonumber
\mathcal{L}^{(b)}_{n,m}&=\left(F^2\right)^{m-1}F_{\mu\nu}F^{\rho\sigma}\left(\frac{n}{2}R\tensor{\left(R^{n-1}\right)}{^{\mu\nu}_{\rho\sigma}}+\frac{1}{4}(n+4-4 m) (3 n-3+4 m)\tensor{\left(R^{n}\right)}{^{\mu\nu}_{\rho\sigma}}\right)\\ 
&-n\left(F^2\right)^{m-1}F_{\alpha\nu}F^{\rho\sigma}\tensor{R}{_{\mu}^{\alpha}}\left((1+2n)\tensor{\left(R^{n-1}\right)}{^{\mu\nu}_{\rho\sigma}}-(n-1)\tensor{R}{^{\beta}_{\rho}}\tensor{\left(R^{n-2}\right)}{^{\mu\nu}_{\beta\sigma}}\right)\, .
\label{eq:lagegqb}
\end{align} 
Note that for $n=0$ we have  $(m-1) \mathcal{L}^{(a)}_{0,m}=\mathcal{L}^{(b)}_{0,m}=(m-1)(3-4m) (F^2)^m$, so that these Lagrangians are well-defined for all integers $n \geq 0$ and $m \geq 1$.  Let us also mention at this point that, in the case $n=m=1$, the existence of Lagrangians with simple magnetic spherically symmetric solutions has been previously noticed in the literature \cite{MuellerHoissen:1988bp,Balakin:2015gpq,Balakin:2016mnn}, although, to the best of our knowledge, generalizations have not been constructed.  

In order to show that the Lagrangians above belong to the EGQ family, we have to check that they become a total derivative when evaluated on the single-function ansatz \req{eq:dsf} with a vector field strength given by \eqref{Fmagnetic}. For that, we evaluate the Lagrangians on the general SSS metric ansatz given by \eqref{eq:sss1}. We get:

\begin{align}
\label{eq:lagegqevsss0}
\mathcal{L}^{(a)}_{n,m}\Big|_{ds^2_{N,f}, F^m}&=\frac{2^{m+n} \psi^{n-1} P^{2m}}{r^{4m}}\bigg[nH-(3n-3+4m) \psi \bigg]\, , \\\nonumber
\mathcal{L}^{(b)}_{n,m}\Big|_{ds^2_{N,f}, F^m}&=\frac{2^{m+n-2} \psi^{n-2} P^{2m}}{r^{4m}}\bigg[n\psi(F+G+2H)+(n+4-4m)(3n-3+4m)\psi^2\\&-2n(1+2n) \psi H+n(n-1)H^2\bigg]\, ,
\label{eq:lagegqevsss}
\end{align}
where, following the notation of \cite{Deser:2005pc}, we have introduced
\begin{align}
\psi&=\frac{1-f}{r^2}\, ,\quad H=-\frac{f N'}{r N}+\frac{1-f-rf'}{r^2}\, ,\\
F&=\frac{-4fN'-2r fN''-3rN' f'-N(2f'+rf'')}{2r N}\, ,\\
G&=\frac{-2r fN''-3rN' f'-N(2f'+rf'')}{2r N}\,.
\end{align}
which represent several components of the Riemann and Ricci tensors. 

\noindent
We can immediately check that both Lagrangians $\mathcal{L}^{(a)}_{n,m}$ and $\mathcal{L}^{(b)}_{n,m}$, defined back at \req{eq:lagegqa} and \req{eq:lagegqb}, belong to the EGQ class. For that, we use the expressions \req{eq:lagegqevsss0} and  \req{eq:lagegqevsss} above and evaluate them on $N=1$. A direct computation shows that
\begin{align}
\label{eq:n1lagegqevsssa}
r^2 \mathcal{L}^{(a)}_{n,m}\Big|_{ds^2_{f}, F^m}&=\frac{d}{dr} \mathcal{I}_{n,m}^{(a)}\, , \\ 
r^2 \mathcal{L}^{(b)}_{n,m}\Big|_{ds^2_{f}, F^m}&=\frac{d}{dr} \mathcal{I}_{n,m}^{(b)}\, , 
\label{eq:n1lagegqevsss}
\end{align}
where
\begin{align}
\mathcal{I}_{n,m}^{(a)}&= 2^{m+n} P^{2m}\frac{\diff }{\diff r}\bigg[ r^{3-4m} \psi^n \bigg]\, ,\\
\mathcal{I}_{n,m}^{(b)}&=2^{m+n-2}  P^{2m}\frac{\diff }{\diff r}\bigg[(-4+2n+
4m) r^{3-4m} \psi^{n}+nr^{4-4m} \psi' \psi^{n-1} \bigg]\, .
\end{align}
Since these Lagrangians are total derivatives, the corresponding Euler-Lagrange equations for the single-function SSS ansatz vanish identically, showing that they are truly EGQ theories. 
\noindent
Actually, we can refine a bit more this statement by noticing that the equation of motion for the metric function $f(r)$ is algebraic, so that both $\mathcal{L}^{(a)}_{n,m}$ and $\mathcal{L}^{(b)}_{n,m}$ are of the quasitopological type. We check this behaviour in the following subsection.

%Task for Angel: check that  these Lagrangians are indeed of the EGQ class. Evaluate them on the single-function SSS ansatz with a magnetic vector field,

%and find the value of their integrals $\mathcal{I}_{n,m}^{(a,b)}$, 
%\begin{equation}
%r^2\mathcal{L}^{(a)}_{n,m}\Big|_{ds^2_{f}}=\frac{d\mathcal{I}_{n,m}^{(a)}}{dr}\, ,\qquad r^2\mathcal{L}^{(b)}_{n,m}\Big|_{ds^2_{f}}=\frac{d\mathcal{I}_{n,m}^{(b)}}{dr}
%\end{equation}

\subsection{Spherically symmetric solutions with magnetic charge }\label{sec:solEQG}
Let us consider the most general extension of the Einstein-Maxwell theory that can be constructed out of the Lagrangians $\mathcal{L}^{(a)}_{n,m}$ and $\mathcal{L}^{(b)}_{n,m}$ defined in \eqref{eq:lagegqa}, \eqref{eq:lagegqb}, \textit{i.e.}, 

%\begin{equation}
%I=\frac{1}{16\pi}\int d^4x\sqrt{|g|}\left[R+\sum_{n=0}^{\infty}\sum_{m=1}^{\infty} \ell^{2(n+m-1)}\left(\lambda_{n,m} \mathcal{L}^{(a)}_{n,m}+\gamma_{n,m}\mathcal{L}^{(b)}_{n,m}\right)\right]\, .
%\label{eq:eqt}
%\end{equation}
\begin{equation}
I=\frac{1}{16\pi}\int d^4x\sqrt{|g|} \mathcal{L}^{\mathrm{EQG}}\, ,    
\label{eq:eqt}
\end{equation}
where
\begin{equation}
\mathcal{L}^{\mathrm{EQG}}=R+\sum_{n=0}^{\infty}\sum_{m=1}^{\infty} \ell^{2(n+m-1)}\left(\lambda_{n,m} \mathcal{L}^{(a)}_{n,m}+\gamma_{n,m}\mathcal{L}^{(b)}_{n,m}\right)\,.
\end{equation}
As explained above, $\mathcal{L}^{(a)}_{0,1}=-F^2$ and $\mathcal{L}^{(b)}_{0,1}=0$, so the usual Maxwell term is included in the action \eqref{eq:eqt} as long as we set $\lambda_{0,1}=1$.  Let us also clarify that we formally include an infinite number of terms in the action out of generality, but at any moment we may consider that only a finite number of couplings are non-vanishing.

Our goal in this section is to obtain magnetically-charged SSS solutions in this set of theories. We have already determined that the Maxwell equation is always solved by magnetic vectors with field strength \eqref{Fmagnetic} and that, due to the properties of the theory, the Einstein's equations allow for solutions with $N=1$. Thus, we only have to determine the equation of the function $f$ in the SSS metric \eqref{eq:sss1}, which can be obtained by evaluating the action on the SSS ansatz \eqref{eq:sss1} (with the help of \req{eq:lagegqevsss}), varying with respect to $N$, and then evaluating at $N=1$. The resulting equation takes the form of a total derivative, $d\hat{\mathcal{E}}(f,r)/dr=0$, and upon integration we obtain the algebraic equation

\begin{equation}
1-f-\frac{2M}{r}+\sum_{n=0}^{\infty}\left(1-f\right)^{n-1}\left[\alpha_{n}(r)+\beta_{n}(r) f\right]=0\, ,
\label{eq:eomf}
\end{equation}
where $M$ is an integration constant that can be straightforwardly related to the mass of the solution. In this expression we have introduced two functions $\alpha_n$ and $\beta_n$ that are given by 
\begin{align}
\alpha_{n}(r)=\sum_{m=1}^{\infty} \frac{\alpha_{n,m}}{r^{4m+2n-2}}\, , \quad
\beta_{n}(r)=\sum_{m=1}^{\infty} \frac{\beta_{n,m}}{r^{4m+2n-2}}
\end{align}
where the coefficients depend on the magnetic charge and on the coupling constants of the theory as 

\begin{align}
 \alpha_{n,m} &=2^{m+n-1} P^{2m} \ell^{2(n+m-1)}\bigg[\lambda_{n,m}+(m-1) \gamma_{n,m}\bigg ]\, ,\\ 
 \beta_{n,m} &=2^{m+n-2} P^{2m} \ell^{2(n+m-1)}\bigg[2(n-1) \lambda_{n,m}+(n^2-4n+2+m(-2+4n)) \gamma_{n,m}\bigg ]\, ,
\end{align}
As we remarked earlier, the equation of motion for $f(r)$ \eqref{eq:eomf} seems to be the most general equation one can get for the set of all Electromagnetic Quasitopological gravities. This is, we suspect that any other EQG will only have the effect of changing the value of the coefficients $\alpha_{n,m}$ and $\beta_{n,m}$ above. We explicitly show in appendix \ref{app:a} that any Quasitopological theory built with two curvature tensors and two gauge field strengths indeed satisfies this property.

Along with the metric and the (magnetic) field strength there is an additional physical magnitude of interest that we can find: the electric potential associated to the dual field strength. Indeed, if $(g_{\mu \nu}, F_{\mu \nu})$ is a magnetic solution of a given theory, then $(g_{\mu \nu}, G_{\mu \nu})$, where $G_{\mu \nu}$ is the dual field strength as defined in Equation \eqref{eq:dualfield}, is a solution of the associated dual theory. In this latter theory, the potential of $G_{\mu \nu}$ will be electric. This electric potential will make its appearance in the subsequent first law of black hole thermodynamics (see Subsection \ref{subsec:bhth}), so it will be useful to compute it. For that, using the notation employed in Eq. \eqref{eq:defpandm}, first we notice that 

%Besides being interesting to study the profile of this dual electric field, the electric potential plays an important role when writing the first law of black hole mechanics. 
%Since it remains invariant under duality transformations, we conclude that the electric potential of a magnetic solution (that is, the potential for the dual field strength) is a physically relevant quantity which will also play a fundamental role in the putative first law of our original Electromagnetic quasi-topological gravity. 
%We will understand better this feature with the explicit calculations presented in Subsection \ref{subsec:bhth}, but for the time being it is clear that it will be useful to compute the electric potential associated to our solutions. 
%For that, using the notation employed at eq. \eqref{eq:defpandm}, first we notice that 
\begin{equation}
-2\mathcal{M}_{\mu \nu}=Y_{\mu \nu \alpha \beta } F^{\alpha \beta} (F^2)^{m-1}+(m-1)Y_{\lambda \eta \alpha \beta } F^{\alpha \beta} F^{\lambda \eta} F_{\mu \nu} (F^2)^{m-2}\, ,
\end{equation}
where we have implicitly defined
\begin{equation}
\begin{split}
Y_{\lambda \eta \alpha \beta}&=\sum_{n=0}^\infty \sum_{m=1}^\infty \lambda_{n,m} \left (  2n \tensor{R}{_{\mu [\alpha}} \tensor{(R^{n-1})}{^{\mu}_ {\beta] \lambda \eta}}+2n \tensor{R}{_{\mu [\lambda}} \tensor{(R^{n-1})}{^\mu_{ \eta]\alpha \beta}}-2(3n-3+4m)(R^n)_{\lambda \eta \alpha \beta} \right )\\&+\sum_{n=0}^\infty \sum_{m=1}^\infty \gamma_{n,m}\left ( n R(R^{n-1})_{\lambda \eta \alpha \beta}+\frac{1}{2}(n+4-4m)(3n-3+4m)(R^n)_{\lambda \eta \alpha \beta}\right. \\&-n(1+2n) \tensor{R}{_{\mu [\alpha}} \tensor{(R^{n-1})}{^{\mu}_ {\beta] \lambda \eta}} -n(1+2n) \tensor{R}{_{\mu [\lambda}} \tensor{(R^{n-1})}{^\mu _{ \eta] \alpha \beta}}\\& \left. +2n(n-1)\tensor{R}{^\mu_{[\lambda}} \tensor{(R^{n-2})}{_{\eta]\mu  \sigma [\alpha}} \tensor{R}{^\sigma_{\beta]}}\right )  \,.
\end{split}
\end{equation}

Imposing the  field strength to be magnetic \eqref{Fmagnetic} and evaluating on the general SSS ansatz \eqref{eq:sss1}, we find that the dual field strength $G$, given by Equation \eqref{eq:dualfield}, takes the form
\begin{equation}
G=\sum_{n=0}^\infty \sum_{m=1}^\infty \ell^{2(n+m-1)} 2^{n+m-3} m P^{2m-1} \mathcal{G}_{n,m} \diff t \wedge \diff r\, ,
\end{equation}
where we defined
\begin{equation}
\begin{split}
\mathcal{G}_{n,m}=-\frac{\psi^{n-2}}{r^{4m-2}}& \left ( 2nr(n \gamma_{n,m}+2\lambda_{n,m}) \psi \psi'+(n-1) n r^2 \gamma_{n,m} (\psi')^2\right .\\&\left.+ \psi(-2(4m-3)((n+2m-2) \gamma_{n,m}+2\lambda_{n,m}) \psi+nr^2 \gamma_{n,m}\psi'') \right )\,.
\end{split}
\end{equation}
Amusingly, the latter can be explicitly rephrased as $G=-\Psi'(r) \diff t \wedge \diff r$, $\Psi$ being the electric potential. It takes the form
\begin{equation}
\Psi =\sum_{n=0}^\infty \sum_{m=1}^\infty \ell^{2(n+m-1)} 2^{n+m-3} m P^{2m-1}  r^{3-4m} \psi^{n}  \bigg[ 4 \lambda_{n,m} +\bigg  ( (-4+2n+4m)+nr\frac{\psi'}{\psi} \bigg )\gamma_{n,m} \bigg] \, .
\label{eq:eqpotential}
\end{equation}
%The SSS solution constructed out of solutions of the equation of motion \eqref{eq:eomf} do not need to represent a black hole solution, since it is not guaranteed that the metric function $f(r)$ will vanish at some value of the radial coordinate. Likewise we shall show in Subsection [[[]]] the general existence of regular solutions for $f(r)$, both in the black hole and non-black hole case. 
Once we have under control all the physically relevant magnitudes, we are going to present next some explicit examples of SSS solutions of particular Electromagnetic Quasitopological gravities, since it will help us illustrate some of the most important features of this new type of theories.

%We will be mainly interested in studying those cases in which the SSS solution is identified with a black hole metric. Therefore, we shall be concerned next in analysing properties of SSS black hole solutions of the most general EGQ theory.

\subsection{Explicit non-singular solutions in quadratic-curvature theories}

In general, an explicit solution of Eq.~\eqref{eq:eomf} in a theory involving an arbitrary number $n$ of Riemann curvature tensors is not available, since we would need to obtain the roots of an $n$th-degree polynomial. Therefore, for the sake of simplicity, let us analyze the solutions of theories that are only of second order in the curvature $(n\le2)$ but including an arbitrary number of gauge field strengths $(m\ge 1)$.
In this case, the equation of motion for the metric function $f(r)$ takes the form
\begin{equation}
1-f-\frac{2M}{r}+\alpha_0(r)+\left[\alpha_{1}(r)+\beta_{1}(r) f\right]+(1-f)\left[\alpha_{2}(r)+\beta_{2}(r) f\right]=0\,.
\label{eq:eomf2}
\end{equation}
where we have taken into account that $\alpha_0(r)=-\beta_0(r)$. Since this is a quadratic polynomial in $f$ one may solve the equation right away to obtain two possible solutions
\begin{equation}\label{eq:fsolEQG}
\begin{split}
f_{\pm}(r)&=\frac{-\alpha_2(r)+\beta_1(r)+\beta_2(r)-1}{2\beta_2(r)}\\& \pm \frac{\sqrt{(-\alpha_2(r)+\beta_2(r)+\beta_1(r)-1)^2+4\beta_2(r) \bigg [ \alpha_2(r)+\alpha_1(r)+\alpha_0(r)-\dfrac{2M}{r}+1\bigg] }}{2\beta_2(r)}\,.
\end{split}
\end{equation}
Now, one can check that the solution $f_{+}$ is asymptotically flat and that it satisfies $f_{+}(r)= 1-2M/r+\ldots$ when $r\rightarrow\infty$. In addition, this solution reduces to the Reissner-Nordstr\"om one in the limit in which the corrections vanish $\ell\rightarrow 0$,
\begin{equation}
\lim_{\ell\rightarrow 0}f_{+}(r)=1-\frac{2M}{r}+\frac{P^2}{r^2}\, ,
\end{equation}
where one has to take into account that $\alpha_{0}\rightarrow P^2/r^2$ while $\alpha_{1}$, $\alpha_{2}$, $\beta_{1}$ and $\beta_{2}$ vanish in that limit.  On the other hand, $f_{-}$ has an exotic asymptotic behaviour, $f_{-}(r)\sim -1/\beta_{2}(r)$, and it does not have an Einstein gravity limit, so we will consider only $f_{+}$ as the physically relevant solution. 

A very remarkable property of these solutions is that, in many cases, the curvature singularity at $r=0$ is regularized by the higher-derivative corrections. In order to analyze the behaviour of the solution near $r=0$, let us assume that we only include terms containing up to  $2m_{c}$ field strengths (so that $1\le m\le m_c$). Then, the functions $\alpha_{i}$ and $\beta_{i}$ read
\begin{align}
\alpha_{n}(r)=\sum_{m=1}^{m_{c}} \frac{\alpha_{n,m}}{r^{4m+2n-2}}\, , \quad
\beta_{n}(r)=\sum_{m=1}^{m_{c}} \frac{\beta_{n,m}}{r^{4m+2n-2}}\, ,
\end{align}
and we can see that, except for certain fine-tuned values for the couplings,  $\alpha_2(r)$ and $\beta_2(r)$ are the dominant terms in the limit  $r \rightarrow 0$. Hence, from \req{eq:fsolEQG} we get
\begin{equation}
 f_{+}(r)=\frac{- \alpha_{2,m_c}+\beta_{2,m_c} + \vert  \alpha_{2,m_c}+\beta_{2,m_c} \vert }{2 \beta_{2,m_c}}+\mathcal{O}(r^2)\quad\text{when}\quad r\rightarrow 0\,.
\end{equation}
Thus, whenever $\alpha_{2,m_c}+\beta_{2,m_c}>0$  and $\beta_{2,mc}\neq 0$ we have $f_{+}(r)\sim 1+\mathcal{O}(r^2)$ near $r=0$, implying that the geometry is regular there. Notice that this is quite remarkable, since we do not need to fine-tune the couplings, only demand that they satisfy a bound. In order for the solution to be globally regular we also have to make sure that there are no other singularities, \textit{e.g.}, the term inside the square root in \req{eq:fsolEQG} should not become negative, but again this is easily achievable (for instance, if all the couplings are positive).  
Let us note that the regularization is also possible if we only include linear curvature terms ($n=1$) ---  in fact this has been previously observed in the literature in the case of $F^2R$ theories \cite{Balakin:2015gpq}, although in that case the couplings must be related in a specific way. 
By analyzing the solutions of Eq.~\req{eq:eomf} near $r=0$, one can see that the regularization at $r=0$ can also be achieved in the general case in which we include terms $F^{2m}R^n$ of arbitrary order. The form of the equation \req{eq:eomf} forces $f(r)\sim 1+\mathcal{O}(r^2)$ near $r=0$ in most of the cases in a quite natural way. Therefore we conclude that, without the need of much tuning on the coupling constants, the magnetically-charged SSS solutions of the EQG theories \req{eq:eqt} are singularity-free.\footnote{See, \textit{e.g.}, Refs.~\cite{AyonBeato:1998ub,AyonBeato:1999ec,AyonBeato:1999rg,Lemos:2011dq,Olmo:2012nx,Olmo:2015axa,Menchon:2017qed,Sert:2015ykz,Cisterna:2020rkc} for other examples of non-singular black holes in different setups. }

\begin{figure}[t]
\centering
\begin{minipage}{0.49\textwidth}
  \centering
\includegraphics[width=\textwidth]{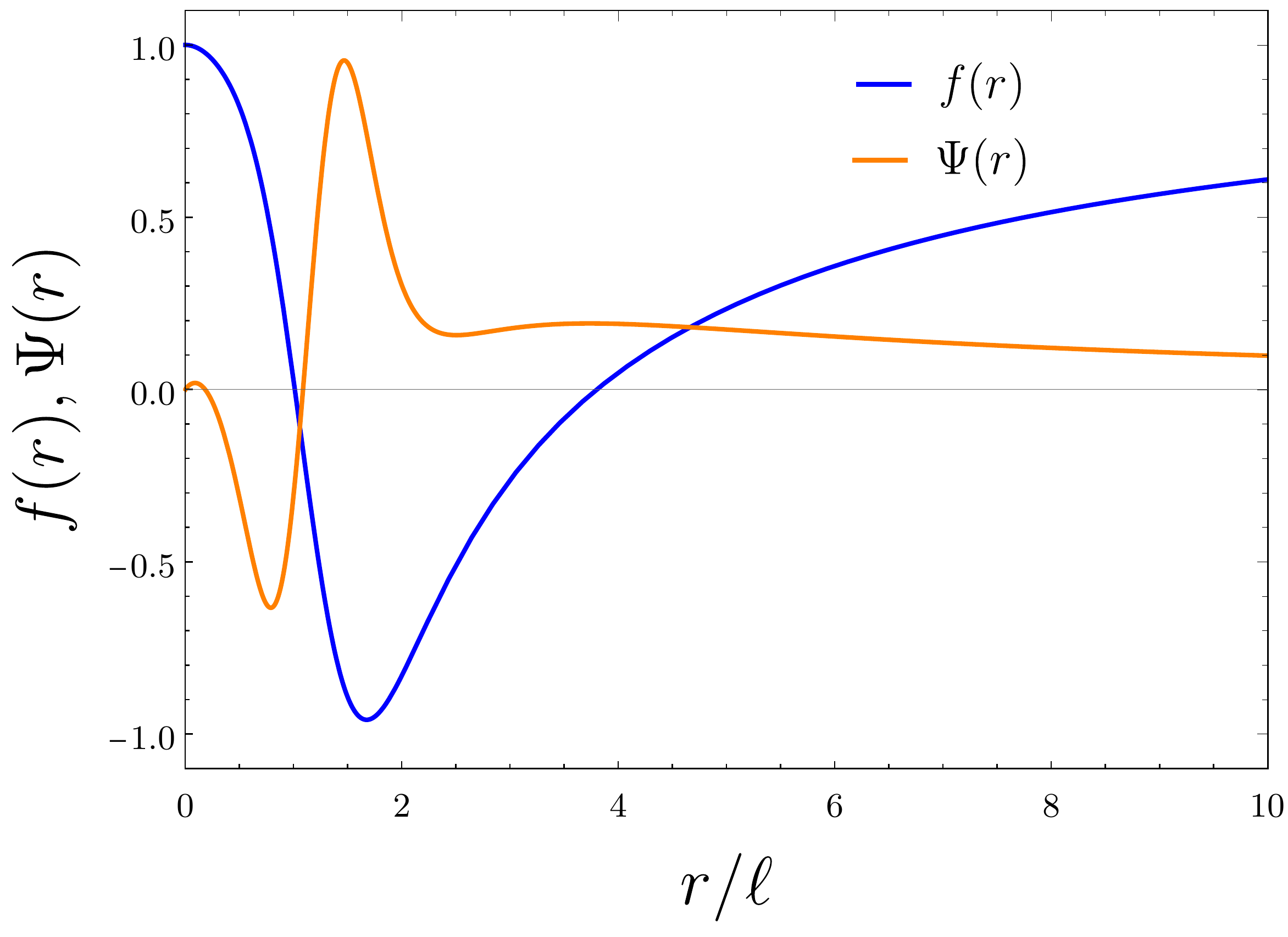}
\subcaption[first caption.]{A black hole solution. We have set $2P=M=2\ell$, $2\gamma_{1,1}=2=-2\gamma_{1,2}=\lambda_{1,1}$ and $\lambda_{1,2}=\frac{25}{16}$.} \label{fig:qta}
\end{minipage}%
\begin{minipage}{0.01\textwidth}
\centering
\hfill
\end{minipage}
\begin{minipage}{0.49\textwidth}
  \centering
\includegraphics[width=\textwidth]{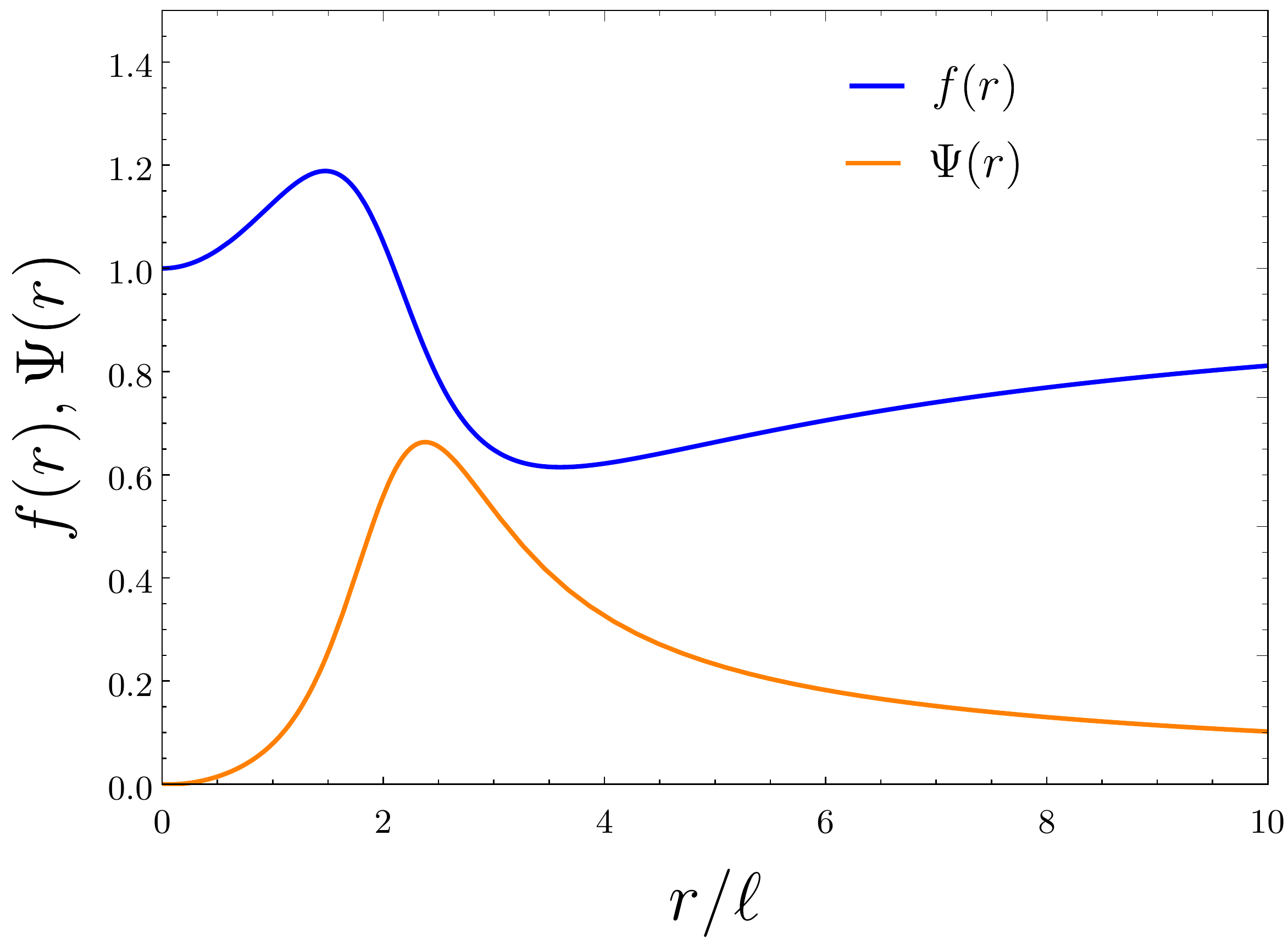}
\subcaption[second caption.]{A non-black hole solution. We have set $P=M=\ell$, $3\gamma_{1,1}=6=-6\gamma_{1,2}=\lambda_{1,1}$ and $\lambda_{1,2}=\frac{53}{4}$.} \label{fig:qtb}
\end{minipage}
\caption{The metric function $f(r)$ and the potential $\Psi$ (in appropriate units) for two given particular sets of couplings, magnetic charge $P$ and mass $M$. Note that Fig. \ref{fig:qta} represents a black hole solution with an inner and outer horizon while Fig. \ref{fig:qtb} is an instance of a non-black hole solution. In both cases, couplings have been chosen so that $\Psi$ is regular at $r=0$, and we see that it vanishes in this limit.} \label{fig:ex1QTG}
\end{figure}

Notice that we have made not reference yet to black hole solutions, because, as in Einstein-Maxwell theory, not all the charged solutions are black holes. If the charge is too large compared to the mass, the solution does not have a horizon and in GR this means that we have a naked singularity. However, in our theories the gravitational field does not diverge, so horizonless regular solutions exist. In Fig~\ref{fig:qta} we show the profile of $f(r)$ for a black hole solution while in Fig.~\ref{fig:qtb} we show the one corresponding to a gravitating point charge. In both cases, the gravitational field is regular everywhere.

Since regular black holes are not possible in GR due to the singularity theorems by Hawking and Penrose, it follows that some of the hypothesis of these theorems are broken by our higher-derivative theories. In particular, these theorems rely on different energy conditions, and these may not be satisfied. Let us check the case of the null energy condition (NEC), which is satisfied if for any future-pointing null vector field $k^{\mu}$ it holds that $T_{\mu\nu}k^{\mu}k^{\nu}\ge0$, where $T_{\mu\nu}$ is the stress-energy tensor, defined ---as if we were working in GR --- by the equation $G_{\mu\nu}=T_{\mu\nu}$. When evaluated on the metric \req{eq:dsf}, it is not difficult to show that, for any null vector $k^{\mu}$ we have
\begin{equation}
T_{\mu\nu}k^{\mu}k^{\nu}=a^2\left(f''(r)+\frac{2(1-f(r))}{r^2}\right)\, ,
\end{equation}
where $a^2$ is a non-negative quantity related to the normalization and direction of the null vector. We have checked that this quantity does in fact become negative for our regular black hole solutions for some intervals of $r$, and hence the NEC is violated. Let us mention anyway that there is no objective way of distinguishing what goes into the right-hand-side or left-hand-side of Einstein's equations in the theory \req{eq:eqt}, so that the definition of the stress-energy tensor is rather arbitrary.

Another physical quantity of interest is the dual electric potential $\Psi$, which was calculated back in Equation \eqref{eq:eqpotential}. In the particular Electromagnetic Quasitopological theories we are considering, it turns out that the electric potential takes the form
\begin{equation}
\begin{split}
\Psi(r)&=\sum_{m=1}^\infty \frac{\alpha_{0,m} m}{P \, r^{4m-3}}+\sum_{m=1}^\infty \frac{m}{P \, r^{4m-1}} (1-f)\left [ (\alpha_{1,m}+\beta_{1,m}) +\frac{1}{2(2m-1)} \beta_{1,m} \frac{-2+2f-rf'}{1-f} \right] \\&+\sum_{m=1}^\infty \frac{m}{P\, r^{4m+1}}(1-f)^2 \left[\alpha_{2,m}+\frac{1}{2m}(\beta_{2,m}-\alpha_{2,m})+\frac{1}{4m} \frac{-2+2f-rf'}{1-f}(\beta_{2,m}-\alpha_{2,m}) \right]\,.
\end{split}
\end{equation}
%where we are defining the coefficients $\alpha_{n,m}$ and $\beta_{n,m}$ as those appearing in the expansions $\alpha_n(r)=\sum_{m=1}^\infty \alpha_{n,m} r^{2-4m-2n}$ and $\beta_n(r)=\sum_{m=1}^\infty \beta_{n,m} r^{2-4m-2n}$.

While the geometry is generally regular, this is not the always case for the electric potential. If we want to have a regular potential at $r=0$, not any set of couplings is allowed. Indeed, since around $r=0$ the metric function $f(r)$ can be approximated by $f(r)\underset{r \sim 0}{\sim} 1+Ar^2$, we have
\begin{equation}
\lim_{r\rightarrow 0} \Psi(r)=\sum_{m=1}^\infty \frac{m}{P\, r^{4m-3}}\left [\alpha_{0,m}-A( \alpha_{1,m}+\beta_{1,m} )-A^2(\alpha_{2,m}+\frac{1}{2m}(\beta_{2,m}-\alpha_{2,m} )) \right ]\,.
\end{equation}
Therefore regularity requires that $\alpha_{0,m}-A( \alpha_{1,m}+\beta_{1,m} )-A^2(\alpha_{2,m}+\frac{1}{2m}(\beta_{2,m}-\alpha_{2,m} ))=0$ for all $m$. This will happen for a certain subset of the whole moduli space of couplings, but it is a realizable feature, as it is shown in Fig. \ref{fig:qta} and Fig. \ref{fig:qtb}. Thus, in these cases black holes and horizonless solutions have regular gravitational and electromagnetic fields everywhere. These solutions are generalizations of the example recently reported in Ref.~\cite{Cano:2020ezi}.

\subsection{Black hole thermodynamics}
\label{subsec:bhth}
After describing the static spherically symmetric solutions of Electromagnetic Quasitopological gravities \eqref{eq:eqt}, in this subsection we focus on black holes and their thermodynamic description. One of our goals is to check that the first law of black hole mechanics holds in these theories and to identify the relevant thermodynamic potentials.

Let us begin with the solution given by \eqref{eq:dsf} and assume the metric function $f(r)$ has some zero for $r \in \mathbb{R}^+$. The black hole horizon would be consequently located at $r_h=\mathrm{max}\{r \in \mathbb{R}^+ \vert f(r)=0\}$. Using the equation of motion for $f(r)$ \eqref{eq:eomf}, after evaluation on $r=r_h$ we find that
\begin{equation}
1-\frac{2M}{r_h}+\sum_{n=0}^\infty \alpha_n(r_h)=0\,.
\end{equation}
From here we can solve for the mass $M$ of the black hole and get
\begin{equation}
2M=r_h+r_h\sum_{n=0}^\infty \alpha_n(r_h)\,.
\label{eq:masseq}
\end{equation}
We can also obtain the temperature, for which we first work out the derivative of \eqref{eq:eomf} at $r=r_h$,
\begin{equation}
-f'(r_h)+\frac{2M}{r_h^2}+\sum_{n=0}^\infty \bigg [-f'(r_h)(n-1) \alpha_n(r_h)+\alpha'_n(r_h)+\beta_n(r_h) f'(r_h) \bigg]=0\,.
\end{equation}
Substituting the expression for the mass found at \eqref{eq:masseq} and taking into account that the temperature $T$ of the black hole is given by $4\pi T=f'(r_h)$, we are left with
\begin{equation}
T=\frac{1}{4\pi r_h }\frac{1+\sum_{n=0}^\infty \bigg[ \alpha_n(r_h)+r_h \alpha'_n(r_h) \bigg]}{1-\sum_{n=0}^\infty\bigg[ \beta_n(r_h)-(n-1)\alpha_n(r_h) \bigg]}\,.
\label{eq:bhtemp}
\end{equation}
Our next objective is the computation of the black hole entropy $S$, which is given by Wald's formula
\begin{equation}
S=-2\pi \int_{\Sigma} d^2x\sqrt{h}\frac{\partial \mathcal{L}}{\partial R_{ \mu \nu\rho\sigma}} \epsilon_{\mu \nu}\epsilon_{\rho \sigma}\, ,
\label{eq:waldentro}
\end{equation}
where $\epsilon_{\mu \nu}$ denotes the binormal to the horizon $\Sigma$. Upon contraction with the binormals, we realize that we just have to care about the component $\dfrac{\partial \mathcal{L}}{\partial R_{ trtr}}$, which turns out to be
\begin{equation}
\frac{\partial \mathcal{L}}{\partial R_{trtr}}\bigg|_{r=r_h}=-\frac{1}{16\pi }\bigg[\frac{1}{2}+\sum_{n=0}^\infty \sum_{m=1}^\infty 2^{n+m-3} n\,  \ell^{2(n+m-1)} \gamma_{n,m} \frac{P^{2m}}{r_h^{4m+2n-2}} \bigg]\,.
\end{equation}
Plugging this result into Eq. \eqref{eq:waldentro} and performing the integration on the angular variables, we find the black hole entropy:

\begin{equation}
S=\pi r_h^2\bigg[1+\sum_{n=0}^\infty \sum_{m=1}^\infty 2^{n+m-2} n\,  \ell^{2(n+m-1)} \gamma_{n,m} \frac{P^{2m}}{r_h^{4m+2n-2}} \bigg]\,.
\label{eq:entropy}
\end{equation}
We see that the entropy is no longer just the black hole area divided by 4 and we have corrections. In particular, we check that the type $(a)$ theory described by \eqref{eq:lagegqa} does not introduce any corrections to the entropy, being just the type $(b)$ theory (and more concretely, the term involving a Ricci scalar) \eqref{eq:lagegqb} the one which causes deviations from the Bekenstein-Hawking result.

In order to check the first law of black hole mechanics we need to bear in mind that we have a magnetically-charged solution. Consequently, if we consider the dual theory, this magnetic solution will become an electric one after dualization. However, we know that electric solutions satisfy a first law which includes the associated electric potential. Therefore, taking into account that black hole thermodynamics remain unchanged under duality transformations, we conclude that the dual of any electric solution, which will be magnetic, will satisfy a first law including the aforementioned electric potential. Hence we can consider this argument backwards to justify that it is the dual electric potential the one entering in the first law of thermodynamics of these magnetic solutions.

From \req{eq:eqpotential} we get the following value of the dual electrostatic potential evaluated at the horizon
\begin{equation}
\Psi_{h}=\sum_{n=0}^\infty \sum_{m=1}^\infty \ell^{2(n+m-1)} m \frac{2^{n+m-1}  P^{2m-1}}{r_h^{4m+2n-3}} \bigg [\lambda_{n,m}+(m-1) \gamma_{n,m}-n\pi r_h T \gamma_{n,m} \bigg]\,,
\label{eq:magpotrh}
\end{equation}
where at the same time we may use the expression for $T$ in \req{eq:bhtemp}.
At this point we have the quantities $M$, $T$, $S$ and $\Psi_{h}$ expressed explicitly as functions of $r_h$ and $P$, and it is not hard to check using \req{eq:masseq}, \req{eq:bhtemp}, \req{eq:entropy} and \req{eq:magpotrh} that the following relations hold,
%If we define the functions $h(r_h,P)$ and $g(r_h,P)$ as follow:
%\begin{equation}
%\begin{split}
%2h(r_h,P)&:=2M=r_h+r_h\sum_{n=0}^\infty \alpha_n(r_h)\,,\\
%g(r_h,P)&:=G_N S=\frac{1}{4}4\pi \bigg[r_h^2+\sum_{n=0}^\infty \sum_{m=1}^\infty 2^{n+m-2} n\,  \ell^{2(n+m-1)} \gamma_{n,m} \frac{P^{2m}}{r_h^{4m+2n-4}} \bigg]\,
%\end{split}
%\end{equation}
\begin{equation}
\begin{split}
T=\frac{\partial_{r_h} M}{\partial_{r_h} S} \, , \quad & \Psi_{h}=\partial_P M-\partial_P S \frac{\partial_{r_h} M}{\partial_{r_h} S}\,,
\end{split}
\label{eq:thermoeasy}
\end{equation}
where $\partial_{r_h}$ and $\partial_{P}$ denote partial differentiation with respect to $r_h$ and $P$, respectively. When expressed this way, one may directly check that the following first law 
\begin{equation}
\diff M=T \diff S+\Psi_{h} \diff P\,,
\label{eq:1stlaw}
\end{equation}
is satisfied. Hence we have shown that there exists a first law of thermodynamics for the Electromagnetic Quasitopological gravities described by the action \eqref{eq:eqt} which holds exactly. Despite the presence of non-minimally coupled terms, this result shows that the first law is formally unchanged, \textit{i.e.}, the effect of the charge appears through the standard term $\Psi_h dP$, where  $\Psi_h$ is the electrostatic potential at the horizon. 

In order to complete our study of the thermodynamic properties of these black holes, let us compute the free energy. On the one hand, this can be defined from the rest of thermodynamic potentials as 
\begin{equation}\label{eq:freeneregyeqg}
F=M-TS\, .
\end{equation}
On the other hand, $F$ should be obtained from the on-shell evaluation of the Euclidean action according to $F=T I^{E}$.
Regarding the computation through the Euclidean action, we have to include an appropriate boundary term and suitable counterterms. Finding these boundary terms for higher-curvature theories of gravity is a highly non-trivial issue, \textit{e.g.} \cite{Teitelboim:1987zz,Myers:1987yn,Deruelle:2009zk,Smolic:2013gz,Teimouri:2016ulk}, but nevertheless one can see that, whatever these terms are, they do not contribute to the on-shell evaluation of the action in the case at hands. On general grounds, we expect that the boundary terms will be proportional to the first derivative of the Lagrangian with respect to the curvature \cite{ECGholo}, and since we are considering asymptotically flat situations, all such terms decay too fast at infinity to make a finite contribution.\footnote{The situation is different for asymptotically AdS solutions, but in that case one may introduce an effective boundary term which is proportional to the Gibbons-Hawking-York term \cite{ECGholo}. This procedure is known to work at least for theories of the GQG class and has been tested in several occasions \cite{Bueno:2018uoy,Bueno:2018yzo,Bueno:2020odt}. }
Thus, we may use as a boundary term the standard Gibbons-Hawking-York term \cite{York:1972sj,Gibbons:1976ue} minus its background contribution. Therefore we propose the following Euclidean action
\begin{equation}
I^E=-\frac{1}{16\pi }\int_M \diff^4 x\, \sqrt{\vert g \vert} \mathcal{L}^{\mathrm{EQG}}-\frac{1}{8\pi} \int_{\partial M} \bigg[ \sqrt{h} K-\sqrt{h^{\mathrm{flat}}} K^{\mathrm{flat}} \bigg ]\,,
\label{eq:eucac}
\end{equation}
where we have already Wick-rotated the time coordinate. 
In order to evaluate this action on our black hole solutions, let us note that the on-shell Lagrangian takes the form of an explicit total derivative when evaluated on the single-function metric \req{eq:dsf}. This follows from \req{eq:n1lagegqevsssa},  \req{eq:n1lagegqevsss} and from a similar property satisfied by the Ricci scalar. Thus we have 
\begin{equation}
\mathcal{L}^{\mathrm{EQG}}\Big|_{ds^2_f, F^m}=\frac{1}{r^2}\frac{d\mathcal{I}}{dr}\, , \quad \mathcal{I}=2r(1-f)-r^2 f'+\sum_{n=0}^\infty \sum_{m=1}^\infty \ell^{2(n+m-1)} \left (\lambda_{n,m} \mathcal{I}_{n,m}^{(a)} +\gamma_{n,m} \mathcal{I}_{n,m}^{(b)} \right )\,.
\end{equation}
and the Euclidean action reads

\begin{equation}
I^E=-\frac{\beta}{4}\mathcal{I}(r)\Big|^{\infty}_{r_h}-\frac{1}{8\pi} \int_{\partial M} \bigg[ \sqrt{h} K-\sqrt{h^{\mathrm{flat}}} K^{\mathrm{flat}} \bigg ]\,,
\end{equation}
Then, one can check that the evaluation of $\mathcal{I}(r)$ at $r\rightarrow\infty$ is exactly cancelled by the boundary terms, so we are left with the evaluation at the horizon, $I^{E}=\beta\mathcal{I}(r_h)/4$. This yields the following value
\begin{equation}
\begin{split}
I^{E}/\beta&=\frac{r_h}{2}+\sum_{n=0}^\infty \sum_{m=1}^\infty \frac{ \ell^{2(n+m-1)} 2^{m+n-2} P^{2m}}{r_h^{4m+2n-3}} \bigg (\lambda_{n,m}+(m-1)\gamma_{n,m} \bigg )\\&-T \bigg ( \pi r^2_h +\sum_{n=0}^\infty \sum_{m=1}^\infty \frac{ \ell^{2(m+n-1)} 2^{m+n-2} P^{2m}}{r_h^{4m+2n-4}} \pi n  \gamma_{n,m} \bigg )\,.
\end{split}
\label{eq:enerlibreeucac}
\end{equation}
By comparison with \req{eq:freeneregyeqg} we check that $F= I^{E}/\beta$, and consequently, we show the inner consistency of  these computations and that the Noether-charge and that the Euclidean path integral approaches to black hole thermodynamics give equivalent results.

Finally, let us work out the specific heat $C_P$ at constant magnetic charge.  A direct application of the inverse function theorem shows that
\begin{equation}
C_P=\left ( \frac{\partial T}{\partial r_h} \right )^{-1}\frac{\partial M}{\partial r_h} =\frac{T (\partial_{r_h} S)^2}{\partial_{r_h}^2 M \, -T \, \partial_{r_h}^2  S}\,.
\label{eq:cpqt}
\end{equation}
We check that $C_P$ generally vanishes in the extremal limit $T \rightarrow 0$, which we study in more detail next.

\subsection{Extremal black holes}\label{sec:EQGextremal}

\begin{figure}[t!]
\centering
\begin{minipage}{0.49\textwidth}
  \centering
\includegraphics[width=\textwidth]{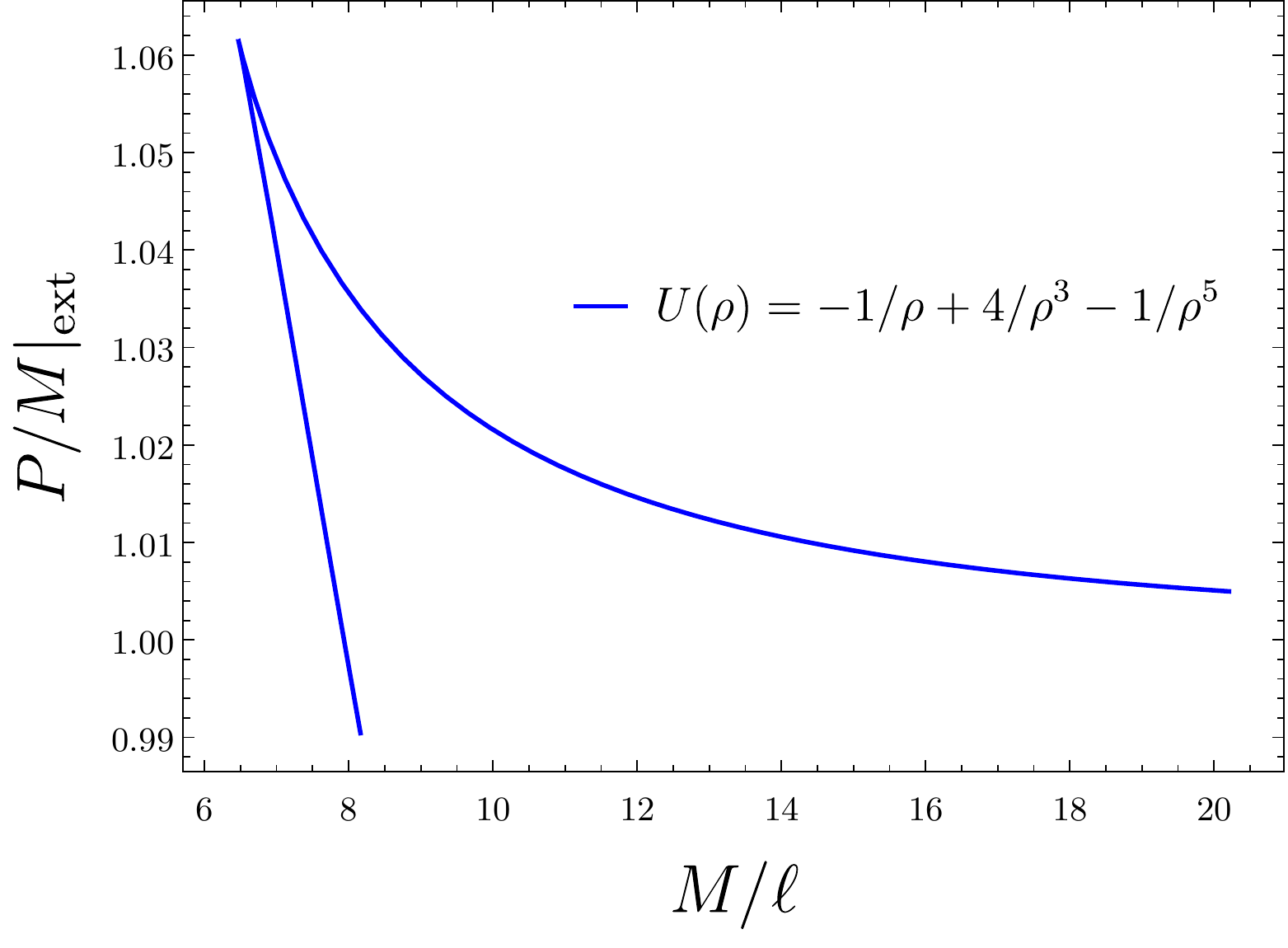}
\subcaption[first caption.]{} \label{fig:ctm-qt1}
\end{minipage}%
\begin{minipage}{0.01\textwidth}
\centering
\hfill
\end{minipage}
\begin{minipage}{0.49\textwidth}
  \centering
\includegraphics[width=\textwidth]{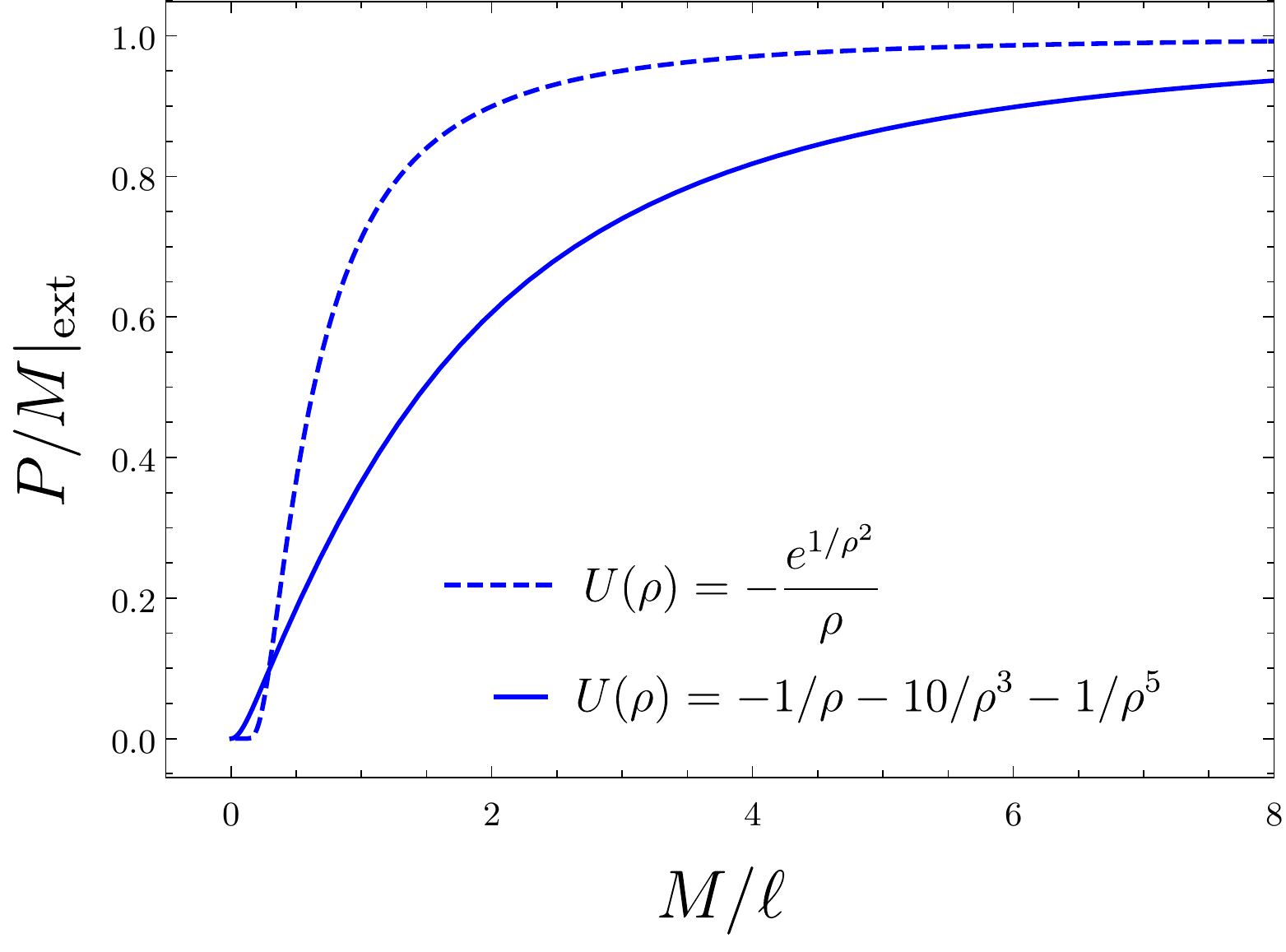}
\subcaption[second caption.]{} \label{fig:ctm-qt2}
\end{minipage}
\caption{The extremal charge-to-mass ratio of particular EQGs as a function of the mass (in units of $\ell$). On the one hand, in Fig. \ref{fig:ctm-qt1} we notice the existence of two branches (just the upper one would be connected to the Reissner-Nordstr\"om black hole), and the charge-to-mass ratio is monotonically decreasing with the mass for both branches. However, there are no extremal black holes below the mass at which both branches merge. On the other hand, in Fig. \ref{fig:ctm-qt2} we consider two different functions $U(\rho)$ and we realize that in both cases the extremal charge-to-mass ratio is monotonically growing.} \label{fig:ctm-qt}
\end{figure}

We finish the study of black holes in Electromagnetic Quasitopological theories by pinpointing some characteristic features of their extremal limit. For simplicity, we restrict ourselves to the particular class of theories which are quadratic in the vector field strength ($m=1$).
If we define the dimensionless parameter $\rho=r_h/\ell$, then we can express the black hole mass $M$ for this class of theories as
\begin{equation}
\frac{2M}{\ell}=\rho-U(\rho) \frac{P^2}{\ell^2}\, ,
\label{eq:eqmasseq}
\end{equation}
where we have introduced the function 
\begin{equation}
U(\rho)=- \sum_{n=0}^\infty \frac{2^n}{\rho^{2n+1}}\lambda_{n,1}\, .
\label{eq:defueq}
\end{equation}
From \req{eq:bhtemp}, we see that the extremality condition $T=0$ consequently takes the form 
\begin{equation}
0=1-\frac{P^2}{\ell^2} U'(\rho)\,.
\label{eq:eqchargeeq}
\end{equation}
From here it is trivial to solve for $P^2$ and obtain the following extremal charge-to-mass ratio:
\begin{equation}
\left. \frac{P}{M} \right \vert_{\rm ext}=\frac{2 \sqrt{U'(\rho)}}{(\rho\,  U'(\rho)-U(\rho) )}\,. 
\end{equation}
Thus, we have an explicit a non-perturbative expression for the extremal charge-to-mass ratio in terms of the radius. Note that if only a finite number of terms is included in the action, the function $U$ is a polynomial in $1/\rho$. However, if an infinite number of them is added, $U$ can actually be any function of the form $U(\rho)=u(\rho^{-2})/\rho$, where $u(x)$ is an arbitrary analytic function (to recover Einstein-Maxwell theory at low energies it must satisfy $u(x)\rightarrow 1$ when $x\rightarrow 0$, though). In Fig.~\ref{fig:ctm-qt} we show  $\left. P/M \right \vert_{\rm ext}$ as a function of the mass for several choices of this function. 

The effect of higher-derivative corrections on extremal black holes has a particular interest in the context of the weak gravity conjecture (WGC) \cite{ArkaniHamed:2006dz}. In fact, a mild form of the WGC states that the extremal charge-to-mass ratio in a consistent theory of quantum gravity must not decrease as the mass decreases. Thus, $P/M\big|_{\rm ext}$ must be a growing (or constant) function  when when we move from larger to smaller masses. This condition ensures that the decay of an extremal black hole into a set of smaller black holes is possible, at least from the point of view of energy and charge conservation. 
Perturbative higher-derivative corrections to the extremal charge-to-mass ratio have been recently explored in a number of papers, \textit{e.g.}  \cite{Cheung:2018cwt,Hamada:2018dde,Bellazzini:2019xts,Charles:2019qqt,Loges:2019jzs,Goon:2019faz,Cano:2019oma,Cano:2019ycn,Andriolo:2020lul,Loges:2020trf}. Our study, on the other hand, is fully non-perturbative, and we can analyze what happens when the corrections become important. 

According to the WGC, just the theory depicted at Fig.~\ref{fig:ctm-qt1} would be admissible, since it satisfies (for the two branches) that the charge-to-mass ratio decreases when the mass grows. However, note that extremal black hole solutions cease to exist below a minimal mass (when the two branches merge). Although this might seem to be a peculiar feature of the particular model considered, this behaviour turns out to appear quite generally. Additional examples of this situation are shown in Section~\ref{sec:extEGQG} for a different family of theories. 
\begin{figure}[t!]
\centering
\includegraphics[width=0.6\textwidth]{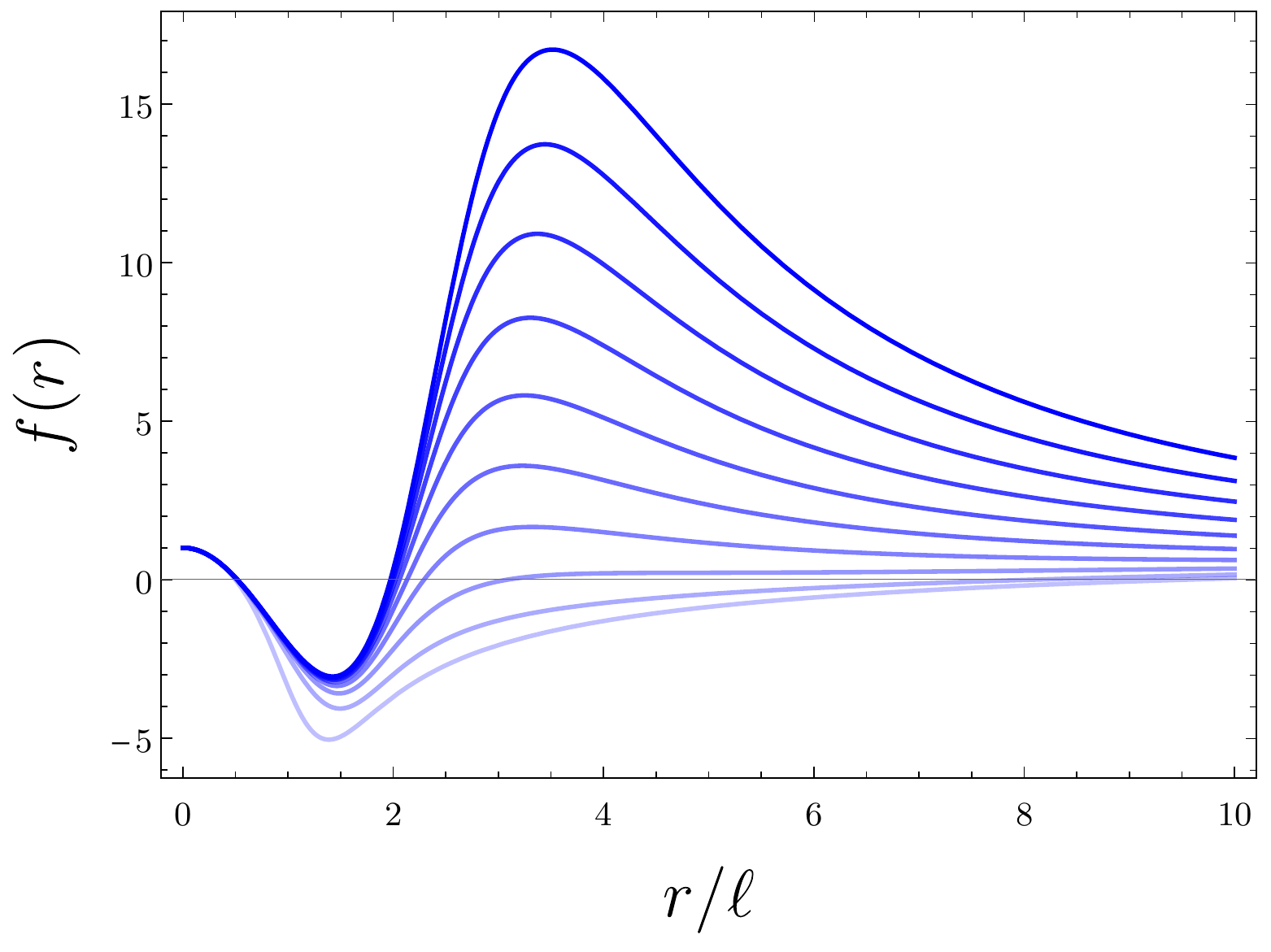}
\caption{Profile of the metric function $f(r)$ for black holes corresponding to the same theory as in Fig~\ref{fig:ctm-qt1}. We show the solutions for a mass $M=5\ell$ (which is below the minimal extremal mass) and, from less to more opacity, $P/\ell=2, 4,\ldots, 20$. As we see, extremality is not reached.} \label{fig:No-Ext}
\end{figure}
One should wonder what happens with the evaporation process of black holes at this point. For that, let us consider an initially large $(M>>\ell)$ non-extremal black hole. Due to Hawking radiation, it loses mass until it approaches extremality. At that moment, it also needs to lose charge in order to continue evaporating, and this is achieved if the WGC holds by emitting a particle with charge-to-mass ratio $p/m\ge 1$. Note that since our black holes satisfy $P/M\big|_{\rm ext}>1$ and this quantity becomes larger for smaller black holes, evaporation is not obstructed. In addition, a process by which an extremal black hole decays into a set of smaller, non-extremal black holes would be in principle allowed in terms of energy and charge conservation. Through this process, the black hole evaporates down to arbitrarily small masses, following approximately the line of extremal black holes in Fig.~\ref{fig:ctm-qt1} (we may assume the black hole remains near-extremal during the evaporation process). Then, it will reach the minimal mass in order for extremal black holes to exist, and this can have several meanings. One possibility is that below that line there are no black holes at all, \textit{e.g.}, all the solutions are naked singularities or horizonless smooth configurations. The black holes could then transition to one of these objects, but this is highly speculative. A more interesting possibility is that below that mass any black hole is non-extremal, and this is precisely the case with the model depicted in Fig.~\ref{fig:ctm-qt1}, corresponding to $\lambda_{1,1}=-2$, $\lambda_{2,1}=1/4$. The black hole solutions of that theory with $M=5\ell$ (below the minimal extremal mass) are shown in Fig.~\ref{fig:No-Ext}. We see that, no matter how large the charge is, the black hole is non-extremal. Thus, in this case, the black hole can always lose mass by means of Hawking radiation without imposing any conditions on kind of particles emitted.  The fact that there is no obstruction to achieve the black hole's evaporation is in fully agreement with the spirit of the WGC.

\begin{figure}[t!]
\centering
\includegraphics[width=0.6\textwidth]{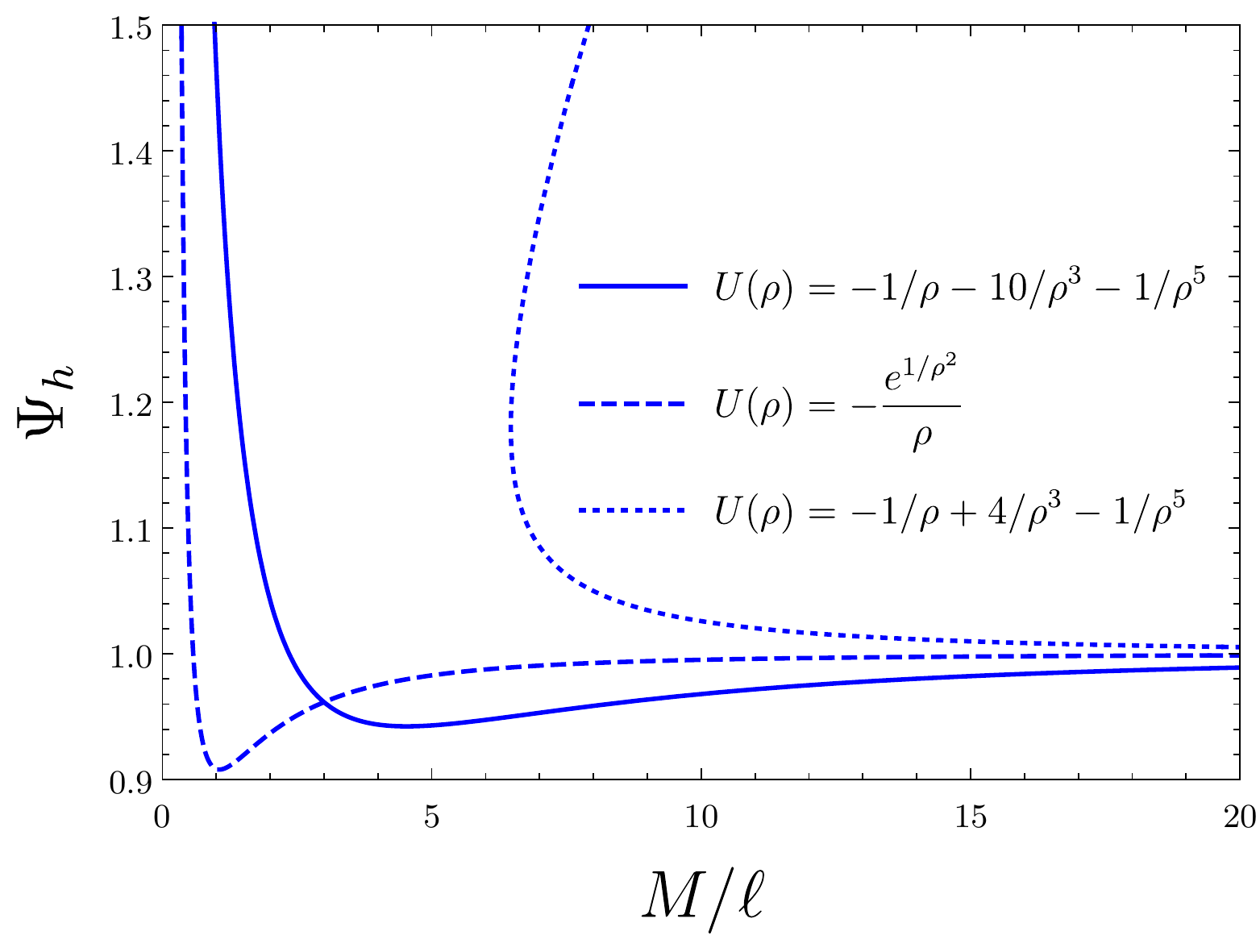}
\caption{The electrostatic potential at the event horizon of extremal black holes as a function of the black hole mass. We plot three different Electromagnetic Quasitopological gravities, specified at the legend of the graph.} \label{fig:potential}
\end{figure}
Finally, let us comment on another interesting property that we can study in the extremal limit, namely, the value of the electrostatic potential $\Psi$ at the event horizon. Indeed, in the limit $T \rightarrow 0$ the quantity $\Psi_{h}$ takes the following surprisingly simple expression: 
\begin{equation}
\Psi_{h}=-\frac{U(\rho)}{\sqrt{U'(\rho)}}\,.
\end{equation}
Let us remark that $\Psi_{h}=1$ for extremal Reissner-Nordstr\"om black holes, but this is no longer the case for our black holes. 
In Fig. \ref{fig:potential} we represent the electrostatic potential for the same three theories we considered in Fig.~\ref{fig:ctm-qt}. We observe that in general such electric potential does not necessarily monotonically increase or decrease with the mass. We discover a rather counter-intuitive fact: rapid decreases in the charge-to-mass ratio plots seem to correspond to increases of the electric potential. Indeed, one would expect that as the charge diminishes, the potential to decrease as well, but we are finding precisely the opposite behaviour. This phenomenon not only happens for the theories forbidden by the WGC, but it also takes place for theories which, a priori, would be allowed. This is explicitly seen for one branch (the one disconnected from Reissner-Nordstr\"om) of the theory with $U(\rho)=-1/\rho+4/\rho^3-1/\rho^5$.

\section{Electromagnetic Generalized Quasitopological gravities}\label{sec:EGQG}
We have just studied two families of EGQGs belonging to the quasitopological subset, \textit{i.e.}, they yield an algebraic equation for the metric function $f$ allowing for the analytic study of the black hole solutions. 
However, these are not the only type of EGQGs that exist. Analogously to the case for pure gravity, there are some theories for which $f$ does not satisfy an algebraic equation, but a 2nd order differential equation --- these are the proper \emph{Generalized} Quasitopological theories. One may wonder why study these theories since we have already described two infinite classes of theories with simpler black hole solutions. There is a good reason, though. It turns out that, even though for Generalized Quasitopological theories we usually are not able to provide an explicit black hole solution, we can, nevertheless, obtain the thermodynamic properties of black holes exactly. 
As we show below, the thermodynamic relations can have a quite different form with respect to the quasitopological case, so that these new theories provide us with qualitatively different modifications of the Reissner-Nordstr\"om solution. 

Based on our previous experience with (purely gravitational) GQGs, we expect that there are many of these theories at each order, so we will not attempt to provide a complete classification of this family of theories. Instead, our goal is to show that these theories indeed exist and to study some of their properties. A general characterization of EGQ theories may be addressed elsewhere. 

We have found a simple family of EGQ Lagrangians, which read
\begin{equation}\label{EGQTGclass}
\mathcal{L}^{\rm \small EGQG}_{n,m}=\left(R^{n-1}\right)^{\mu\nu}\left[nRg^{\alpha\beta}-(4n+4m-3)R^{\alpha\beta}\right]F_{\mu\alpha}F_{\nu\beta}\left(F^2\right)^{m-1}\, ,
\end{equation}
where $\left(R^{n}\right)^{\mu\nu}$ is the $n$-power of the Ricci tensor,
\begin{equation}
\tensor{\left(R^{n}\right)}{^{\mu}_{\nu}}=\tensor{R}{^{\mu}_{\alpha_1}}\tensor{R}{^{\alpha_1}_{\alpha_2}}\ldots \tensor{R}{^{\alpha_{n-1}}_{\nu}}\, ,
\end{equation}
with the convention that $\tensor{\left(R^{0}\right)}{^{\mu}_{\nu}}=\tensor{\delta}{^{\mu}_{\nu}}$. Let us first of all show that these Lagrangians indeed satisfy the GQG condition, as given by Eq.~\req{EGQGcond}. Evaluation on the the general magnetic SSS ansatz,
\begin{align}
\diff s_{N,f}^2=&-N^2(r) f(r) \diff t^2+\frac{\diff r^2}{f(r)}+r^2 \left(\diff \theta^2+\sin^2\theta\diff \phi^2\right)\, ,\\
F=&P d\theta\sin\theta\wedge d\phi\, ,
\end{align}
yields the following reduced Lagrangian,
\begin{equation}\label{eq:redEGQT}
\mathcal{L}^{\rm \small EGQG}_{n,m}\Big|_{\diff s_{N,f}^2}=\left(\frac{2 P^2}{r^4}\right)^m H^{n-1}\left[n R-(4n+4m-3)H\right]\, ,
\end{equation}
where 
\begin{align}\label{eq:Rvalue}
R=&-f''-\frac{4 f'}{r}-\frac{2 f}{r^2}+\frac{2}{r^2}-\frac{N'}{N}\left(3f'+\frac{4 f }{r}\right)-\frac{2 fN''}{N}\, ,\\
\label{eq:Hvalue}
H=&\frac{1-f-rf'}{r^2}-\frac{N'f}{Nr}\, .
\end{align}
This expression will be useful to compute the equations of motion later. Further evaluation on $N=1$ shows that the Lagrangian becomes a total derivative,
\begin{equation}\label{eq:EGQTint}
\mathcal{L}^{\rm \small EGQG}_{n,m}\Big|_{\diff s_{1,f}^2}=\frac{1}{r^2}\frac{d}{dr}\mathcal{I}_{n,m}\, ,\quad \mathcal{I}_{n,m}=\left(\frac{2 P^2}{r^4}\right)^m r^3\left(\frac{1-f-rf'}{r^2}\right)^n\, .
\end{equation}
and therefore it belongs to the EGQG class. Let us now study the black hole solutions of these theories.

\subsection{Black holes}
Let us consider an extension of Einstein-Maxwell theory with the terms of the EGQ class \req{EGQTGclass} above:
\begin{equation}\label{EGQTGaction}
I=\frac{1}{16\pi}\int d^4x\sqrt{|g|}\left[R+\sum_{n=0}^{\infty}\sum_{m=1}^{\infty}\ell^{2(n+m-1)}\mu_{n,m}\mathcal{L}^{\rm \small EGQG}_{n,m}\right]
\end{equation}
Here $\ell$ is an overall length scale while $\mu_{n,m}$ are dimensionless couplings. Note that the usual Maxwell term $-F^2$ is included in the sum, since we have $\mathcal{L}^{\rm \small EGQG}_{0,m}=-(4m-3)(F^2)^m$. Thus, by convention we set $\mu_{0,1}=1$. By construction, this theory has magnetically-charged solutions of the form \req{eq:dsf}, and hence we only have to determine the equation of motion for $f$. This can be done most easily by taking the variation of the reduced Lagrangian $L_{N,f}$ with respect to the function $N$ and evaluating at $N=1$. The result is a third-order equation for $f$ which takes the form of a total derivative: 
\begin{equation}
\frac{d}{dr}\mathcal{G}_{f}=0\, ,
\end{equation}
where 
\begin{align}
\notag
\mathcal{G}_{f}&=r(1-f)+\sum_{n,m}\mu_{n,m}\frac{\ell^{2(n+m-1)}}{2r}  \left(\frac{2P^2}{r^4}\right)^m \left(\frac{1-f-rf'}{r^2}\right)^{n-2} \Big[-(n-1) r^2 f'^2+(n-2) r f'\\
&+f \left((n-1) n r^2 f''+r (2-4 m n) f'+4 m n+2 n^2-3 n-2\right)+f^2 \left((3-4 m) n-2 n^2+1\right)+1\Big]\, .
\label{eqfGQTG}
\end{align}
Note that all the terms in the sum are of second order in derivatives except for those with $n=0$ and $n=1$, since in those cases the Lagrangians reduce to minimally coupled terms $(F^2)^m$ and to Electromagnetic Quasitopological theories as the ones studied in the  Section \ref{sec:EQG}, respectively.  Now, integrating the equation above we get
\begin{equation}\label{eq:eqGQTG}
\mathcal{G}_{f}=2M\, ,
\end{equation}
where $M$ is an integration constant that, as we show next, turns out to be the mass. Since in general the equation above is a differential equation of second order, there are two more integration constants which need to be fixed by the boundary conditions. This is analogous to the case of purely gravitational GQ theories which has been studied in various papers \cite{PabloPablo2,Hennigar:2017ego,PabloPablo4,Ahmed:2017jod}, so let us just comment briefly on it. 

First, we impose that the solution is asymptotically flat (we do not have a cosmological constant), which implies that $f(r)\rightarrow1$ when $r\rightarrow\infty$. In the asymptotic region, we may expand the general solution in the following form:
\begin{equation}\label{fpfh}
f(r)=f_p(r)+f_h(r)\, ,
\end{equation}
where $f_p$ is a particular solution while $f_h$ represents a deviation with respect to that solution (and will satisfy a homogeneous equation). We can obtain a particular solution by  assuming a $1/r$ expansion, which yields the following result
\begin{equation}
f_p(r)=1-\frac{2M}{r}+\frac{P^2}{r^2}+3\mu_{1,1}\frac{\ell^2 P^2}{r^4}+\mathcal{O}(r^{-5})\, .
\end{equation}
On the other hand, the boundary conditions imply that $f_h\rightarrow 0$ asymptotically, and hence we can assume that it is arbitrarily small. Thus, plugging \req{fpfh} into \req{eq:eqGQTG} and expanding linearly in $f_h$ we get
\begin{equation}
a f_h''+b f_h'+c f_h=0\, ,
\end{equation}
where the asymptotic expansion of the coefficient reads
\begin{equation}
a=\frac{2\mu_{2,1}\ell^4P^2}{r^3}+\mathcal{O}\left(\frac{1}{r^4}\right)\, ,\quad b=-\frac{6\mu_{2,1}\ell^4P^2}{r^4}+\mathcal{O}\left(\frac{1}{r^5}\right)\, ,\quad c=-r+\mathcal{O}\left(\frac{1}{r^3}\right)\, .
\end{equation}
The equation above can be solved in terms of Bessel functions, but for our purposes it suffices to note that the asymptotic solution behaves as
\begin{equation}
f_h(r)\sim A \exp\left[{\frac{r^3}{3P\ell^2\sqrt{2\mu_{2,1}}}}\right]+B \exp\left[-{\frac{r^3}{3P\ell^2\sqrt{2\mu_{2,1}}}}\right]\, ,
\end{equation}
where $A$ and $B$ are integration constants. 
Thus, when $\mu_{2,1}>0$ one of the modes is exponentially growing and the other one is exponentially decaying. By setting the appropriate constant to $0$ we achieve an asymptotically flat solution with a free integration constant. When $\mu_{2,1}<0$, the solutions become highly oscillating at infinity and the only way to obtain a regular solution is to set $A=B=0$, thus there are no further boundary conditions that one can fix. This is problematic because we cannot impose regularity at the horizon (see below), and therefore there are no regular black hole solutions in this case. Thus, we only consider $\mu_{2,1}>0$. If this coefficient is $0$, the constraint will appear in the next coefficient in the expansion. 

On the other hand, we impose the existence of a regular horizon, \textit{i.e.}, a point $r_h$ at which $f(r_h)=0$ and around which $f$ is analytic. In particular, we assume that $f$ has a Taylor expansion of the form
\begin{equation}\label{eq:frhexp}
f(r)=4\pi T(r-r_h)+\sum_{n=2}^{\infty}a_n (r-r_h)^n\, ,
\end{equation}
where we are making explicit that $f'(r_h)=4\pi T$, where $T$ is Hawking's temperature. When we insert this expansion into the equation \req{eq:eqGQTG}, we get a system of equations that relate the coefficients $a_n$. Nonetheless, the first two equations are special, since they only involve $r_h$ and $T$. These read
\begin{align}
\label{MGQTG}
M=&\frac{r_h}{2}\left[1+\frac{1}{2}\sum_{n,m}\mu_{n,m}\left(\frac{\ell^2(1-4\pi T r_h)}{r_h^2}\right)^{n-1}\left(\frac{2\ell^2P^2}{r_h^4}\right)^m\left(1+(n-1)4\pi T r_h\right)\right]\, ,\\\notag
0=&1-4\pi T r_h+\frac{1}{2}\sum_{n,m}\mu_{n,m}\left(\frac{\ell^2(1-4\pi T r_h)}{r_h^2}\right)^{n-1}\left(\frac{2\ell^2P^2}{r_h^4}\right)^m\Big(3-4m-2n\\
&+(n+4m-3)4\pi T r_h\Big)\, .
\label{TGQTG}
\end{align}
These two equations allow one to get (implicitly) the temperature $T$ and the radius $r_h$ once $M$ and $P$ are given. The rest of the equations provide relations for the coefficients $a_n$. A simple inspection reveals the only free parameter in the expansion is $a_2$, and the rest of the $a_n$ are fixed in terms of it. 
Finally, $a_2$ is fixed by demanding that the solution be asymptotically flat. The full solution $f(r)$ can be obtained by a numeric integration of \req{eq:eqGQTG} using \req{eq:frhexp} as initial condition, and implementing a shooting algorithm to search for the value of $a_2$ that yields the correct asymptotic behaviour.  Such numeric resolution will be carried out elsewhere, but comparing with previous works on neutral black holes in GQ theories, we expect that the solution exists providing the condition on the couplings discussed above is satisfied. Fortunately, a great deal of information about these black holes can be obtained without resorting to the numeric solution.

\subsection{Black hole thermodynamics}\label{sec:thermoEGQG}
Even though the profile of the solutions has to be determined numerically, one remarkable property of the theories in Eq.~\req{EGQTGaction}  --- which is shared by all the theories of the GQ class --- is that the thermodynamic properties of black holes can be found analytically. First, note that the two relations \req{MGQTG}, \req{TGQTG} above give us the relation between $M$, $P$ and $T$. Unfortunately, such relation cannot be written explicitly due to the complicated form of the equations, but it is nevertheless  possible to solve the system of equations parametrically. Let us first introduce two dimensionless parameters $p$ and $x$ defined as
\begin{equation}\label{psdef}
p=\frac{2\ell^2P^2}{r_h^4}\, ,\qquad x=\frac{\ell^2(1-4\pi T r_h)}{r_h^2}\, .
\end{equation}
Then, we can define the 2-variable function 
\begin{equation}\label{Wfunction}
\mathcal{W}(x,p)=\frac{1}{2}\sum_{n=0}^{\infty}\sum_{m=1}^{\infty}\mu_{n,m}x^np^m\, .
\end{equation}
In terms of these quantities we can rewrite \req{MGQTG} and \req{TGQTG} as follows
\begin{align}\label{eq:Massps}
M=&\, \frac{r_h}{2}\left[1+\left(1-\frac{xr_h^2}{\ell^2}\right)\partial_x\mathcal{W}+\frac{r_h^2}{\ell^2}\mathcal{W}\right]\, ,\\
&&\notag\\\label{rhsp}
0=&\, r_h^2\left(x\partial_x\mathcal{W}+4p\partial_p\mathcal{W}-3\mathcal{W}-x\right)+\ell^2\partial_x\mathcal{W}\, .
\end{align}
Whenever $\partial_x\mathcal{W}\neq 0$, we can obtain explicitly $r_h(x,p)$ from the second equation. On the other hand, if $\mathcal{W}$ does not depend on $x$, the same equation determines the relation $x(p)$, while $r_h$ is free. Note that this only happens in the trivial case in which the higher-order Lagrangians do not depend on the curvature, and hence it is not relevant for our purposes. 
Then, inserting $r_h(x,p)$ in \req{eq:Massps} we obtain the explicit relation $M(x,p)$, and  analogously, we get $T(x,p)$ and $P(x,p)$ from \req{psdef}, namely,

\begin{equation}\label{PTxt}
P=\frac{r_h^2}{\ell}\sqrt{\frac{p}{2}}\, ,\qquad T=\frac{1}{4\pi r_h}\left(1-\frac{x r_h^2}{\ell^2}\right)\, .
\end{equation}
Thus, we have been able to write all these thermodynamic quantities, as well as the radius, in terms of two independent parameters $x$ and $p$. This is a useful way to study the thermodynamic phase space of these theories. 
Let us now compute the rest of thermodynamic properties of these black holes.

The entropy is computed by Wald's formula, as introduced previously in Section~\ref{sec:EQG}
\begin{equation}
S=-2\pi\int d^2x\sqrt{h}\frac{\partial\mathcal{L}}{\partial R_{\mu\nu\rho\sigma}}\epsilon_{\mu\nu}\epsilon_{\rho\sigma}\, .
\end{equation}
When computing the derivative with respect to the curvature of the Lagrangians \req{EGQTGclass}, each time we derive one of the Ricci tensors $R_{\alpha\beta}$ we end up generating a contraction between the binormal $\epsilon_{\alpha\beta}$ and a field strength. Note that such contractions are always $0$ for magnetic configurations, since $\epsilon$ and $F$ are orthogonal in that case. Thus, only the action of the derivative on the Ricci scalar appearing in \req{EGQTGclass} yields a non-vanishing contribution. The result reads
\begin{equation}
S=\pi r_h^2\left[1+\sum_{n,m}\mu_{n,m} n\left(R^{n-1}\right)^{\mu\nu}F_{\mu\alpha}\tensor{F}{_{\nu}^{\alpha}}\left(F^2\right)^{m-1}\right]\Bigg|_{r=r_h}\, ,
\end{equation}
where we have already performed the integration on the horizon. Evaluating this expression at $r=r_h$ and using the parameters \req{psdef} and the function \req{Wfunction}, we get the simple result

\begin{equation}\label{Sxt}
S=\pi r_h^2\left[1+2\partial_x\mathcal{W}\right]\, .
\end{equation}

\noindent
Again, using \req{rhsp} we obtain the explicit relation $S(x,p)$. 

Let us now compute the electrostatic potential at the horizon. As we saw in Section~\ref{sec:dual}, the dual field strength is given by \req{eq:dualfield}. The derivative of our Lagrangians \req{EGQTGclass} with respect to the field strength yields
\begin{equation}
\frac{\partial\mathcal{L}_{n,m}}{\partial F^{\mu\nu}}=2(m-1)F_{\mu\nu}(F^2)^{m-2}F_{\rho\alpha}F_{\sigma\beta}Z^{\rho\sigma\alpha\beta}+2(F^2)^{m-1} F^{\alpha\beta}Z_{[\mu|\alpha|\nu]\beta}\, ,
\end{equation}
where
\begin{equation}
Z_{\rho\sigma\alpha\beta}=\left(R^{n-1}\right)_{\rho\sigma}\left(nR g_{\alpha\beta}-(4n+4m-3)R_{\alpha\beta}\right)\, .
\end{equation}
Evaluating this expression for a magnetic vector field \req{Fmagnetic} and for the metric \req{eq:dsf} we obtain the value of the dual field strength,

\begin{equation}
G=\diff t\wedge \diff r\sum_{n,m}\mu_{n,m} m \frac{P}{r^2}\left(\frac{2\ell^2P^2}{r^4}\right)^{m-1}\ell^{2n}H^{n-1}\left(-nR+(4n+4m-3)H\right)\, ,
\end{equation}
where $R$ and $H$ are given by \req{eq:Rvalue} and \req{eq:Hvalue}. Remarkably enough, this expression takes the form of an explicit total derivative, namely $G=-\Psi'(r) \diff t \wedge \diff r$, where the electrostatic potential reads
\begin{equation}
\Psi(r)=\sum_{n,m}\mu_{n,m} m \frac{P}{r}\left(\frac{2\ell^2P^2}{r^4}\right)^{m-1}\left(\frac{\ell^{2}(1-f-rf')}{r^2}\right)^n\, .
\end{equation}
Finally, evaluating at the horizon and using  \req{psdef} and \req{Wfunction} we may write the result as

\begin{equation}\label{phixt}
\Psi_h=\frac{r_h}{\ell} \sqrt{2p}\partial_p\mathcal{W}\, .
\end{equation}

Additionally, we can obtain the free energy from the on-shell Euclidean action. The computation can be done using the same prescription for the boundary terms as in Eq.~\req{eq:eucac}.  The bulk action can be evaluated right away thanks to our on-shell Lagrangians being total derivatives \req{eq:EGQTint}. Then, the evaluation at infinity gets canceled with the contribution from the boundary terms and we are left with the evaluation of the quantities $\mathcal{I}_{n,m}$ in \req{eq:EGQTint} at the horizon. This yields the following result for the free energy, $F=I_{E}/\beta$:
\begin{equation}\label{Fxt}
F=\frac{r_h}{4}\left(1+x \frac{r_h^2}{\ell^2}+2\mathcal{W}\frac{r_h^2}{\ell^2}\right)\, .
\end{equation}

Summarizing, the equations \req{eq:Massps}, \req{PTxt}, \req{Sxt}, \req{phixt} and \req{Fxt}, together with the relation \req{rhsp}, give us explicit expressions for all the thermodynamic quantities $M$, $P$, $T$, $S$, $\Psi_h$, $F$ and the radius $r_h$ in terms of two independent parameters $x$ and $p$. The theory-dependence of all these formulas is encoded in the function $\mathcal{W}$ defined in \req{Wfunction}. Let us now check that these quantities satisfy consistent thermodynamic relations. In particular, they should satisfy the 1st law of black hole mechanics, 
\begin{equation}
dM=TdS+\Psi_h dP\, .
\end{equation}
This relation is in fact verified. The easiest way to see this consists in assuming first that $r_h$ is an independent variable in the expressions of $M$, $S$ and $P$ (Eqs.~\req{eq:Massps}, \req{Sxt} and \req{PTxt}, respectively). Then, the variations of those quantities with respect to \emph{just} $x$ and $p$ automatically satisfy the first law above. Afterwards, we may take the variation only with respect to $r_h$ (assuming that now $x$ and $p$ are independent variables) and we check that it also satisfies the first law  once we notice the constraint \req{rhsp}. Hence when the dependence of $r_h$ on $x$ and $p$ is taken into account, the first law  holds too for arbitrary variations of the free parameters. 

On the other hand one can also check that $F=M-TS$, which is a non-trivial consistency test of our results, indicating that the Wald's entropy (Noether charge) and the Euclidean action approaches are equivalent. 

\subsection{Extremal and near-extremal black holes}\label{sec:extEGQG}
Let us study how the corrections affect extremal black holes. In terms of the variable $x$, the extremality condition $T=0$ implies that $x$ and $r_h$ are related according to
\begin{equation}
x=\frac{\ell^2}{r_h^2}\, .
\end{equation}
Due to this, \req{rhsp} becomes a complicated equation that determines the relation between $p$ and $x$ at extremality. Namely, we have 
\begin{equation}\label{extremalconst}
2x\partial_x\mathcal{W}+4p\partial_p\mathcal{W}-3\mathcal{W}-x=0\, .
\end{equation}
To simplify the discussion, let us consider the subset of theories that are only quadratic in the Maxwell field strength (but which have an arbitrary number of higher-curvature terms). In such case, the function $\mathcal{W}$ has the form
\begin{equation}
\mathcal{W}=\frac{p}{2}U(x)\, ,\quad\text{where}\quad U(x)=1+\sum_{n=1}^{\infty}\mu_{n,1}x^n\, .
\label{eq:defueqg}
\end{equation}
For this function, it is possible to solve \req{extremalconst} explicitly to obtain $p(x)$ at extremality:
\begin{equation}
p=\frac{2x}{2x U'+U}\, .
\end{equation}
Then, from \req{eq:Massps} and \req{PTxt} we obtain the mass and the charge\footnote{We stumble upon the following fact: the results for the extremal mass and charge given by Eq. \eqref{eq:mpextegq} coincide exactly with those for EQs --- that one may obtain from Eqs. \eqref{eq:eqmasseq} and \eqref{eq:eqchargeeq} --- after performing the replacement $U(x)\rightarrow -\rho\, U(\rho)$, where $U(x)$ and $U (\rho)$ are given by Eqs. \eqref{eq:defueqg} and \eqref{eq:defueq} respectively and where $x=\frac{1}{\rho^2}$. Hence the associated extremal charge-to-mass ratios exhibit the same types of phenomena in both sets of theories.}

\begin{equation}
M_{\rm ext}=\frac{\ell}{\sqrt{x}}\left[\frac{x U'+U}{2xU'+U}\right]\, ,\qquad P_{\rm ext}=\frac{\ell}{\sqrt{x}}\frac{1}{\sqrt{2x U'+U}}\, ,
\label{eq:mpextegq}
\end{equation}
and the extremal charge-to-mass ratio

\begin{equation}
\frac{P}{M}\bigg|_{\rm ext}=\frac{\sqrt{2x U'+U}}{xU'+U}\, .
\end{equation}
The entropy in turn reads

\begin{equation}
S_{\rm ext}=\frac{\pi \ell^2}{x}\left[\frac{4xU'+U}{2x U'+U}\right]\, .
\end{equation}
Then, as we did in Section~\ref{sec:EQGextremal} we can check some particular cases to see if it is possible to satisfy the mild form of the Weak Gravity Conjecture at a non-perturbative level. 

\begin{figure}[t]
	\begin{center}
		\includegraphics[width=0.6\textwidth]{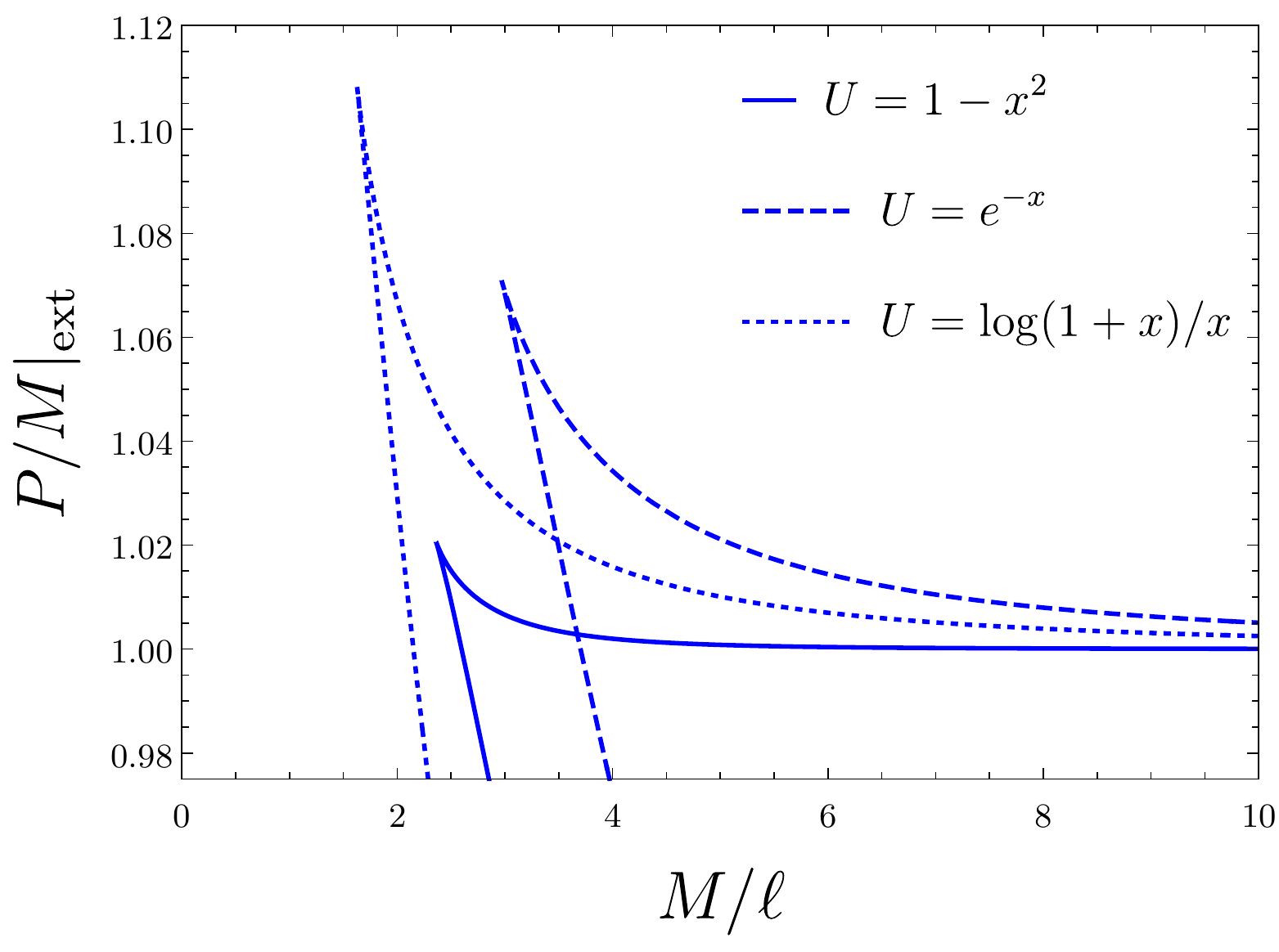} 
		\caption{Extremal charge-to-mass ratio for some higher-derivative theories. The couplings of the higher-derivative terms are chosen so that $P/M$ increases when $M$ decreases, but we see it is not possible to continue this trend all the way down to $M=0$. There is a minimum mass below which extremal black holes do not exist. In these examples we see that each curve has two branches, but only the upper one is smoothly connected with the Reissner-Nordstr\"om solution when the higher-derivative couplings are set to $0$.}
		
		\label{fig:CTMGQTG}
	\end{center}
\end{figure}

Since $x=\ell^2/r_h^2$, we must demand that $P/M\big|_{\rm ext}$ is monotonically growing with $x$, although in that case we also have to make sure that $M$ is a decreasing function of $x$. As an example, let us consider the case in which there is a single higher-derivative term in the action so that  $U=1+\mu_n x^n$. Then we have
\begin{equation}
M_{\rm ext}=\frac{\ell}{\sqrt{x}}\left[\frac{1+(n+1)\mu_nx^n}{1+(2n+1)\mu_nx^n}\right]\,, \quad \frac{P}{M}\bigg|_{\rm ext}=\frac{\sqrt{1+(2n+1)\mu_nx^n}}{1+(n+1)\mu_n x^n}\, ,
\end{equation}
For $\mu_n>0$, the extremal charge-to-mass ratio and the mass are actually monotonically decreasing with $x$, so this case should be discarded according to the WGC. On the other hand, if we take $\mu_n<0$ we observe that for small $x$ (large $M$) the charge-to-mass ratio is in fact growing with $x$. However, it soon reaches a maximum value and then decreases again. Moreover, $M$ has a minimum value, so there are no extremal black holes below certain mass --- see Fig.~\ref{fig:CTMGQTG}.  One can also consider other choices of higher-derivative terms that yield different forms of the function $U(x)$, and a few examples are shown in Fig.~\ref{fig:CTMGQTG}.  We find the same qualitative behaviour in all of these cases, namely, $P/M$ has a maximum value which happens for the minimum mass. Thus, it seems quite difficult for $P/M\big|_{\rm ext}$ to be a growing function all the way down to $M=0$, at least within this family of theories. Nevertheless, this can be interesting from the point of view of the WGC, since, as we saw in Section~\ref{sec:EQGextremal}, it may imply that below the minimal mass all the solutions are non-extremal black holes, and hence there is no obstacle to prevent the evaporation of these black holes. 

Another intriguing fact about these examples is that the corrections to the extremal entropy are negative. 
For instance, in the case of $U=1+\mu_n x^n$ we have 
\begin{equation}
S_{\rm ext}=\frac{\pi \ell^2}{x}\left[\frac{1+(4n+1)\mu_n x^n}{1+(2n+1)\mu_nx^n}\right]< \pi P^2\quad\text{if}\quad\mu_n<0.
\end{equation}
This seems in contradiction with some claims and results in the literature \cite{Cheung:2018cwt,Goon:2019faz} which relate positive corrections to the extremal charge-to-mass ratio with positive corrections to the entropy. However, the contradiction is not such, since, as noted in Ref.~\cite{Cano:2019ycn}, the comparison must be done with the corrections to the near extremal entropy, while the corrections to the extremal entropy are independent. This is an example of that situation. 

\begin{figure}[t]
	\begin{center}
		\includegraphics[width=0.6\textwidth]{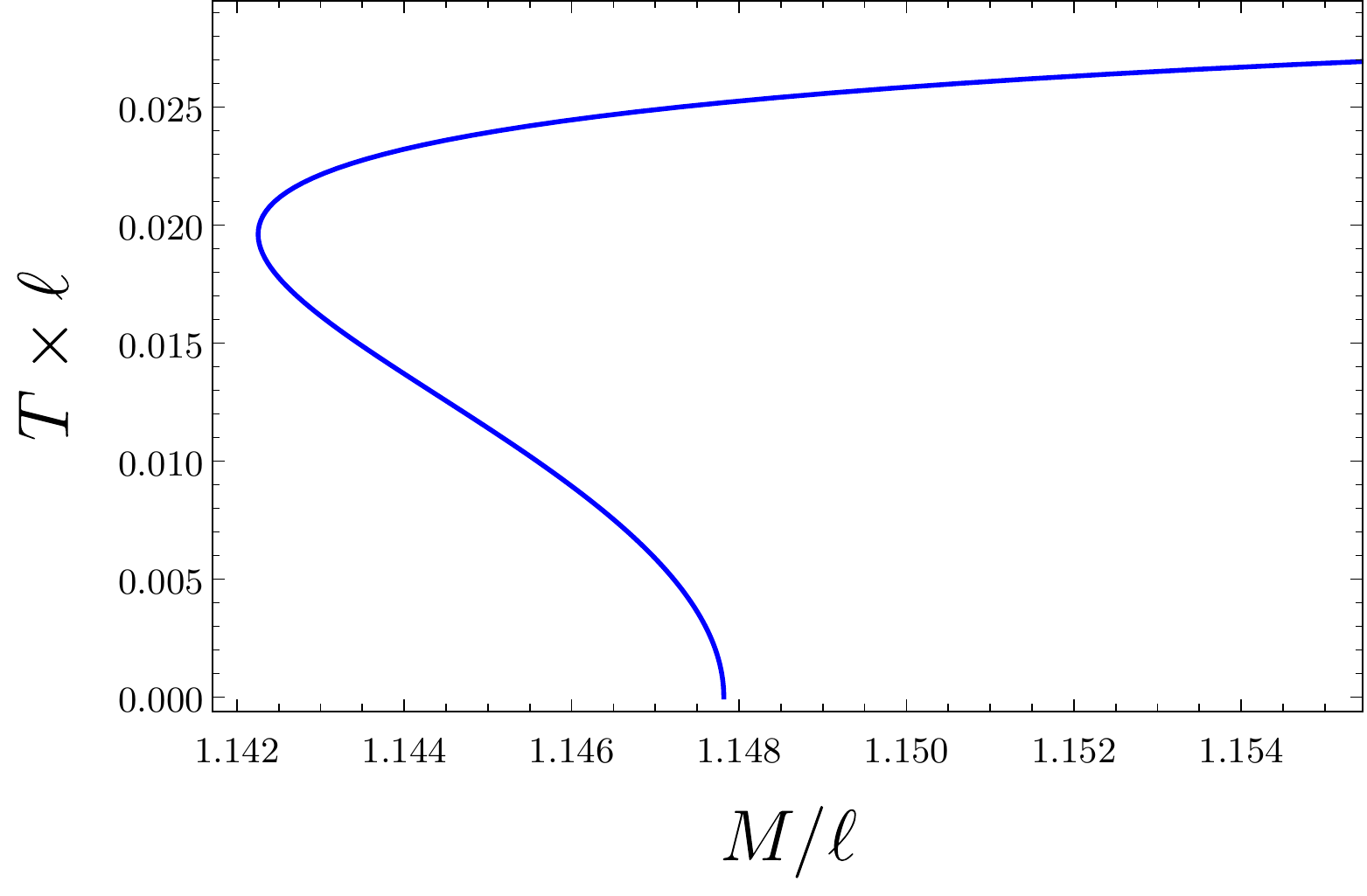} 
		\caption{Temperature vs mass diagram at fixed charge for near-extremal black holes with $\left(\partial^2M/\partial T^2\right)_P\big|_{\rm ext}<0$. The extremal black hole is not the state with the minimal mass. We consider the model $U(x)=1+x^2$, but nevertheless the profile of the curve will be similar for any other case in which $\left(\partial^2M/\partial T^2\right)_P\big|_{\rm ext}<0$.}
		\label{fig:NearExtGQTG}
	\end{center}
\end{figure}

Finally, let us also briefly comment on near-extremal black holes. A characteristic property of extremal black holes in Einstein gravity is that the specific heat at constant charge goes to zero, while its first derivative is positive. This means that near-extremal black holes satisfy $M-M_{\rm ext}=cT^2$ with $c>0$, and therefore the mass of the black hole grows as we increase the temperature. Interestingly enough, this is not always the case for our black holes with higher-derivative corrections. The specific heat, defined as 
$C_P=\left(\frac{\partial M}{\partial T}\right)_P$, vanishes at extremality, but its first derivative reads instead

\begin{equation}
\left(\frac{\partial^2 M}{\partial T^2}\right)_P\bigg|_{\rm ext}=\frac{4 \ell^3\pi ^2 \left(-6 x^2 U U''-10 x^3 U' U''+U^2\right)}{x^{3/2} \left(2 x U'+U\right) \left(x \left(2 x U''+5 U'\right)+U\right)}\, .
\end{equation}
One can see that this quantity can have either sign, depending on the model and on the value of $x$. If it is positive, then near-extremal black holes behave as in Einstein-Maxwell theory and  they are stable, in the sense that when we increase the temperature (hence we depart from extremality) the mass also increases. The case in which this quantity is negative is quite intriguing. It implies that in order to get away from extremality, the black hole must lose mass. Therefore, extremal black holes are thermodynamically unstable and they do not represent the minimal mass state for a given charge. Instead, the minimal mass state will take place at a different point in which $C_P=0$, and this is the solution to which the black hole tends when it evaporates. 
An example of this situation is represented in Fig.~\ref{fig:NearExtGQTG}, where we show $T$ vs $M$ at fixed charge for a particular set of higher-derivative terms. 
Another consequence of this effect is that, in a certain region of the parameter space, there exists more than one black hole solution with the same mass and charge. This non-uniqueness of solutions can be thought as a discrete violation of the no-hair conjecture and  is analogous to the situation with charged black holes in Einsteinian cubic gravity that was recently reported in Ref.~\cite{Frassino:2020zuv}. 

\section{Conclusions}\label{sec:con}
In this paper, we have introduced a new class of non-minimally coupled higher-derivative extensions of Einstein-Maxwell theory. These theories are characterized by possessing magnetic SSS solutions characterized by a single metric function $f$ (see \req{eq:dsf})  whose equation of motion is (at least partially) integrable. In addition, within this set of theories, the thermodynamic properties of black holes can be computed exactly. Such theories are analogous to the Generalized Quasitopological gravities and thus we refer to them as  Electromagnetic Generalized Quasitopological gravities. As in the case of pure gravity, we have seen EGQGs come in two main classes: those for which the SSS equations of motion can be reduced to an algebraic equation for $f$ belong to the ``quasitopological'' class, while if the equation is of second order we say that the theory is properly of the ``generalized quasitopological'' class. We have constructed an infinite number of densities of both types, although we suspect that there are many others, especially in the case of Generalized Quasitopological theories.  Determining the most general structure of these Lagrangians would be an interesting problem.

In the case of Quasitopological theories, we have shown some explicit examples of black hole and non-black hole solutions --- see Section~\ref{sec:solEQG}. We observed that, in a quite remarkable and general way, these solutions possess globally regular geometries, \textit{i.e.}, the timelike singularity at $r=0$ characteristic of charged black holes or point charges is smoothed away by the higher-derivative corrections. In slightly more restrictive cases, we showed that the electrostatic potential of the dual theory also remains finite everywhere, thus making these solutions particularly appealing. In particular, in the horizonless case, one may regard these objects as solitons or even as four-dimensional fuzzballs. 
 
For both, Quasitopological and Generalized Quasitopological theories, we have performed a detailed study of black hole thermodynamics --- see Sections \ref{subsec:bhth} and \ref{sec:thermoEGQG}. We have been able to provide explicit expressions for all the relevant thermodynamic potentials and we have shown that the first law of black hole mechanics, 
\begin{equation}
dM=TdS+\Psi_h dP\, ,
\end{equation}
holds exactly. Here $S$ is Wald's entropy and $\Psi_h$ is the electrostatic potential of the dual theory evaluated on the event horizon. Thus, the first law is formally unchanged with respect to the case of a minimally coupled gauge field. This is not a general proof of the first law, but rather a check of it for a large class of theories. It would be interesting to actually attempt a proof in the case of general $\mathcal{L}(R_{\mu\nu\rho\sigma},F_{\alpha\beta})$ theories, as done in Ref.~\cite{Rasheed:1997ns} in the case of non-linear electromagnetism coupled to Einstein gravity. In addition, we have checked that the Euclidean methods provide the same answer for black hole thermodynamics than the Noether's charge approach. In particular, we have seen that the on-shell Euclidean action yields indeed the free energy, $T I^{E}=F=M-T S$. Due to the large space of theories that we consider, we have not made a general analysis of the features of the new thermodynamic relations, so a more detailed study is left for future work. 

Motivated by the weak gravity conjecture, we did study the properties of extremal and near-extremal black holes in these theories. A mild form of the WGC states that, in a consistent quantum theory of gravity, the charge-to-mass ratio of extremal black holes should grow monotonically as the mass decreases. This would allow for the decay of extremal black holes in terms of energy and charge conservation. Previous literature had studied perturbative corrections to the extremality bound in a variety of theories, ranging from general EFTs to Stringy effective actions \cite{Cheung:2018cwt,Hamada:2018dde,Bellazzini:2019xts,Charles:2019qqt,Loges:2019jzs,Goon:2019faz,Cano:2019oma,Cano:2019ycn,Andriolo:2020lul,Loges:2020trf}. Although our theories do not belong (a priori) to those categories, they have the advantage of allowing us to perform exact, non-perturbative computations. Thus, they may be used to learn about the corrections to extremality at large coupling. As we observed in Sections \ref{sec:EQGextremal} and \ref{sec:extEGQG}, it is always easy (\textit{e.g.}, by choosing the signs of the couplings appropriately) to get $P/M\big|_{\rm ext}$ to satisfy the WGC when the mass is large (\textit{i.e.}, in the perturbative regime). However, when the curve $P/M\big|_{\rm ext}$ vs $M_{\rm ext}$ is continued to lower masses, one often finds that it stops at a minimal mass, meaning that there are no extremal black holes below that mass. There can be different reasons for this behaviour, but we have shown with an example (see Fig.~\ref{fig:No-Ext}) that a possibility is that below that mass all solutions are non-extremal black holes, regardless the value of the charge. This is actually appealing from the point of view of the WGC, since it implies that, below the minimal mass, charged black holes find no obstruction to evaporate. 

Higher-derivative corrections can introduce new effects into the game, and in particular we observed another situation that has implications for black hole evaporation. In some instances --- as shown in Fig.~\ref{fig:NearExtGQTG} --- it may occur that the extremal black hole is not the one with a minimum mass for a given charge. In those cases, (near-)extremal black holes are unstable, and tend to decay to this minimum mass black hole, which has a non-vanishing temperature. Thus, in that situation one does not have to worry about the charge-to-mass ratio of the extremal black hole, but about the one of the minimum mass black hole. An analogous example has been recently reported in Ref.~\cite{Frassino:2020zuv} in the context of Einsteinian cubic gravity with a (minimally-coupled) Maxwell field. 

The new theories offer various possibilities since they allow us to perform many explicit computations that are inaccessible in general higher-derivative theories. Thus, let us close the paper by commenting on future directions. 
As we have already mentioned, it would be interesting to complete the characterization of EGQ Lagrangians to find the most general action of this type. On the other hand, here we have focused on asymptotically flat solutions, so one could extend this work by including a non-vanishing cosmological constant. The asymptotically anti-de Sitter case is particularly relevant due to its connection to holography. In fact, it is known that higher-derivative gravities with a negative cosmological constant are very useful holographic toy models that can be used to learn non-trivial information about Conformal Field Theories --- see Refs.~\cite{Bueno:2018yzo,Bueno:2020odt} for recent results involving Generalized Quasitopological gravities. Since EGQGs contain higher-derivatives not only of the metric but also of a vector field, these may be used to probe additional aspects of a CFT.

One could also characterize subsets of these theories satisfying additional properties. For instance, we expect that some of the EGQGs allow for single-function Taub-NUT solutions whose thermodynamic properties can be studied exactly, as the ones in \cite{Bueno:2018uoy}. In particular, some theories of the quasitopological subclass might allow for explicit Taub-NUT solutions.  In addition, non-minimally coupled electrodynamics may be of interest in cosmology --- see \textit{e.g.} \cite{Bamba:2008ja} --- so we may  wonder if some of these theories could be useful in that context, as the ones in \cite{Arciniega:2018fxj,Cisterna:2018tgx,Arciniega:2018tnn}.

Regarding higher-dimensional generalizations, we find several possibilities. Since the defining property of EGQGs is related to the structure of their magnetic SSS solutions, a straightforward generalization can be achieved in the case of gravity coupled to a $(D-3)$-form in $D$ dimensions. In that case we ask for the same property to hold, namely, that there are magnetic SSS solutions with $g_{tt}g_{rr}=-1$. Note that the dual of these theories corresponds to gravity coupled to a vector field, and the magnetic solutions become electric ones, so they are more natural in this dual frame. 
On the other hand, in higher-dimensions there are solutions in the form of extended objects, such as black strings. One may search for  higher-derivative theories in which the structure of these solutions remains simple, in a way analogously to the condition $g_{tt}g_{rr}=-1$. As an example, one may consider the case of a six-dimensional theory with a non-minimally coupled 2-form field, and search for magnetic black string solutions satisfying such type of condition. These higher-dimensional generalizations may be addressed elsewhere.

\acknowledgments
We would like to thank Pablo Bueno, Tom\'as Ort\'in, Pedro F. Ram\'irez and Carlos S. Shahbazi for useful comments and discussions. 
The work of PAC is supported by the C16/16/005 grant of the KU Leuven. The work of \'AM is funded by the Spanish FPU Grant No. FPU17/04964. \'AM was further supported by the MCIU/AEI/FEDER UE grant PGC2018-095205-B-I00 and by the ``Centro de Excelencia Severo Ochoa'' Program grant SEV-2016-0597.

\appendix

\section{Dualizing theories of quadratic order in $F$}\label{app:dualEGQG}

Let us consider a theory of gravity coupled to electromagnetism given by the following action:
\begin{equation}
I=\frac{1}{16\pi} \int d^4 x \sqrt{\vert g \vert} \mathcal{L}(R_{\mu \nu \rho \sigma}, F_{\alpha \beta}) =\frac{1}{16\pi} \int d^4 x \sqrt{\vert g \vert} \left[ R- \mathcal{Q}_{\mu \nu \rho \sigma}(g,R) F^{\mu \nu} F^{\rho \sigma} \right ]\,,
\label{eq:appaction}
\end{equation}
where  $\mathcal{Q}_{\mu \nu \rho \sigma}$ depends exclusively on the metric $g_{\mu \nu}$ and its associated Riemann curvature tensor $R_{\mu \nu \rho \sigma}$. Note the following symmetry properties of $\mathcal{Q}_{\mu \nu \rho \sigma}$:
\begin{equation}
\mathcal{Q}_{\mu \nu \rho \sigma}=-\mathcal{Q}_{\nu \mu \rho \sigma}=-\mathcal{Q}_{\mu \nu \sigma \rho}=\mathcal{Q}_{\rho \sigma \mu \nu}\,.
\end{equation}
From \eqref{eq:appaction}, one can easily compute that
\begin{equation}
\mathcal{M}_{\mu \nu}=-\frac{1}{2} \frac{\partial \mathcal{L}}{\partial F^{\mu \nu}}=\mathcal{Q}_{\mu \nu \rho \sigma} F^{\rho \sigma}\,. 
\end{equation}
Taking into account Eq. \eqref{eq:dualfield}, we have that 
\begin{equation}
G_{\mu \nu}=(\star \mathcal{M})_{\mu \nu}=\frac{1}{2} \varepsilon_{\mu \nu \alpha \beta} \mathcal{Q}^{\alpha \beta \rho \sigma} F_{\rho \sigma}\,.
\label{eq:appgdual}
\end{equation} 
Therefore the action $I_{\mathrm{dual}}$ dual to \eqref{eq:appaction} turns out to be
\begin{equation}
I_{\mathrm{dual}}=\frac{1}{16\pi}\int d^4 x \sqrt{\vert g \vert} \left[R+ \mathcal{M}_{\mu \nu} F^{\mu \nu} \right ]\, ,
\end{equation}
where we have included the contribution from the Lagrange multiplier term which imposes the Bianchi of $F_{\mu \nu}$. Define now a tensor $\mathcal{Q}^{-1}_{\mu \nu \rho \sigma}$ with the same symmetries as $\mathcal{Q}_{\mu \nu \rho \sigma}$ which satisfies
\begin{equation}
\mathcal{Q}^{-1}_{\mu \nu \rho \sigma} \mathcal{Q}^{\rho \sigma \alpha \beta}=\tensor{\delta}{_{\mu \nu}^{\alpha \beta}}\,.
\end{equation}
Using this inverse tensor of $\mathcal{Q}$, it is clear that
\begin{equation}
F_{\mu \nu}=\mathcal{Q}^{-1}_{\mu \nu \rho \sigma} \mathcal{M}^{\rho \sigma}\, .
\end{equation}
Therefore, on taking into account Eq. \eqref{eq:appgdual}, we find that the dual theory may be expressed in the following compact form:
\begin{equation}
I_{\mathrm{dual}}=\frac{1}{16\pi}\int d^4 x \sqrt{\vert g \vert} \left[ R- \chi_{\mu \nu \rho \sigma} G ^{\mu \nu}  G^{\rho \sigma} \right ]\,,
\label{eq:appdualth}
\end{equation}
where we have defined the tensor $\chi_{\mu \nu \rho \sigma}$ as
\begin{equation}
\chi_{\mu \nu \rho \sigma}=-\frac{1}{4} \varepsilon_{\mu \nu \alpha \beta}(\mathcal{Q}^{-1})^{\alpha \beta \lambda \eta}  \varepsilon_{\rho \sigma \lambda \eta}=6 \delta_{\mu \nu [\rho \sigma} \tensor{\mathcal{Q}}{^{-1}_{\alpha \beta]}^{\alpha \beta}} \,. 
\end{equation}
This procedure can be used to obtain electrically-charged solutions from E(G)Qs with magnetic ones, as it was explicitly done in Ref.~\cite{Cano:2020ezi}. 

In general, the dual theory \eqref{eq:appdualth} will not be polynomial. However, it is possible to write it as a formal power series if we decompose $\mathcal{Q}$ as
\begin{equation}
\mathcal{Q}_{\mu \nu \rho \sigma}=\delta_{\mu \nu \rho \sigma}+\tilde{\mathcal{Q}}_{\mu \nu \rho \sigma}\,.
\end{equation}  
This decomposition is rather natural since it corresponds to an expansion around the pure Maxwell term given by $\delta_{\mu \nu \rho \sigma}$. The inverse $\mathcal{Q}^{-1}$ can in turn be written as
\begin{equation}
\mathcal{Q}^{-1}_{\mu \nu \rho \sigma}=\sum_{n=0}^\infty (-\tilde{\mathcal{Q}})^n_{\mu \nu\rho\sigma}\, ,
\end{equation}
where we have defined  $(\tilde{\mathcal{Q}}^0)^{\mu \nu \rho \sigma}=\delta^{\mu \nu \rho \sigma}$ and $(\tilde{\mathcal{Q}}^n)^{\mu \nu \rho \sigma}=\tensor{\tilde{\mathcal{Q}}}{_{\mu \nu }^{\mu_2 \nu_2}} \tensor{\tilde{\mathcal{Q}}}{_{\mu_2 \nu_2 }^{\mu_3 \nu_3}}\cdots \tensor{\tilde{\mathcal{Q}}}{_{\mu_n \nu_n }^{\rho \sigma}}$. Substituting back in Eq. \eqref{eq:appdualth}, we obtain the dual action in terms of $\mathcal{\tilde{Q}}$:
\begin{equation}
I_{\mathrm{dual}}=\frac{1}{16\pi }\int d^4 x \sqrt{\vert g \vert} \left[ R-G_{\mu \nu} G^{\mu \nu}-6 \sum_{n=1}^\infty  G_{[\rho \sigma}(-\tensor{\tilde{\mathcal{Q}})}{^n_{\alpha \beta]}^{\alpha \beta}}  G^{\rho \sigma} \right ]\,.
\end{equation}

\section{Equations of motion from the reduced action}\label{app:redaction}

In this appendix we show the validity of the reduced Lagrangian method in order to obtain the equations of motion. Let us first of all consider the Lagrangian evaluated on the general SSS metric ansatz \req{eq:sss1} and on either an electric or magnetic vector ansatz. 
Taking into account that the gravitational equations are given by the variation of the action,
\begin{equation}
\mathcal{E}_{\mu\nu}=\frac{1}{\sqrt{|g|}}\frac{\delta I}{\delta g^{\mu\nu}}\, ,
\end{equation}
if $I_{N,f}$ denotes the action one obtains after evaluation on the general SSS ansatz \req{eq:sss1}, we find after using the chain rule

\begin{align}
\mathcal{E}_{N}=&\frac{\delta I_{N,f}}{\delta N}=\sqrt{|g|}\mathcal{E}_{\mu\nu}\frac{\partial g^{\mu\nu}}{\partial N}=\sqrt{|g|}\frac{2}{N^3f}\mathcal{E}_{tt}\, ,\\
\mathcal{E}_{f}=&\frac{\delta I_{N,f}}{\delta f}=\sqrt{|g|}\mathcal{E}_{\mu\nu}\frac{\partial g^{\mu\nu}}{\partial f}=\sqrt{|g|}\left[\frac{1}{N^2f^2}\mathcal{E}_{tt}+\mathcal{E}_{rr}\right]\, .
\end{align}
Then, we see that indeed $\mathcal{E}_{N}=0$ and $\mathcal{E}_{f}=0$ imply $\mathcal{E}_{tt}=\mathcal{E}_{rr}=0$. Next, let us check that all the off-diagonal components of the gravitational equations are trivial. The easiest way to show this is to take into account that the tensor $\mathcal{E}_{\mu\nu}$ satisfies the same symmetries as the metric. Thus we must have
\begin{equation}
L_{k_{(i)}} \mathcal{E}_{\mu\nu}=0, 
\end{equation}
where $L_{k_{(i)}}$ denotes the Lie derivative along any of the $i=1,...,4$ Killing vectors associated to the spacetime symmetries. In terms of the usual spherical coordinates in which the most general SSS metric \eqref{eq:sss1} is expressed, these Killing vectors read as follows:
\begin{equation}
k_{(1)}=\partial_t\, , \quad k_{(2)}=\partial_{\phi}\, , \quad k_{(3)}=-\sin \phi \, \partial_\theta-\cos \phi \cot \theta \, \partial_\phi\, , \quad k_{(4)}= \cos \phi \, \partial_\theta-\sin\phi \cot \theta \, \partial_\phi\,.
\end{equation}
On the one hand, both $L_{k_{(1)}} \mathcal{E}_{\mu\nu}=L_{k_{(2)}} \mathcal{E}_{\mu\nu}=0$ directly imply that all components of $\mathcal{E}_{\mu\nu}$ are independent of both time $t$ and azimuthal coordinate $\phi$. Computing now $\cos \phi L_{k_{(3)}} \mathcal{E}_{\mu\nu}+\sin \phi L_{k_{(4)}} \mathcal{E}_{\mu\nu}$, we find that 
\begin{equation}
\bigg (\cos \phi L_{k_{(3)}} \mathcal{E}_{\mu\nu}+\sin \phi L_{k_{(4)}} \mathcal{E}_{\mu\nu} \bigg )=\begin{pmatrix}
0 & 0 & \csc^2 \theta \mathcal{E}_{t\phi} & -\mathcal{E}_{t\theta} \\
0 & 0 & \csc^2 \theta \mathcal{E}_{r\phi} & -\mathcal{E}_{r\theta} \\
 \csc^2 \theta \mathcal{E}_{t\phi} &  \csc^2 \theta \mathcal{E}_{r\phi} & 2\csc^2 \theta \mathcal{E}_{\theta \phi} & -\mathcal{E}_{\theta \theta}+\csc^2 \theta \mathcal{E}_{\phi \phi} \\
  -\mathcal{E}_{t\theta}  &-\mathcal{E}_{r\theta}& -\mathcal{E}_{\theta \theta}+\csc^2 \theta \mathcal{E}_{\phi \phi} & -2\mathcal{E}_{\theta \phi}\\
\end{pmatrix}\,.
\end{equation}
Since the latter must be identically zero, we learn that  $\mathcal{E}_{\theta \theta}=\csc^2\theta \mathcal{E}_{\phi \phi}$ and that all off-diagonal components of $\mathcal{E}_{\mu\nu}$, except for $\mathcal{E}_{tr}$, vanish. In order to show that $\mathcal{E}_{tr}$ also vanishes, we can perform a direct computation using \eqref{eq:EinsteinEq}.

The term proportional to the metric in \eqref{eq:EinsteinEq} is trivially diagonal, as well as that corresponding to $\tensor{\mathcal{M}}{_{(\mu}^\alpha} F_{\nu) \alpha}$, after taking into account that $\mathcal{M}_{\mu \nu}$ has necessarily the same components as $F_{\mu \nu}$.
Indeed, both antisymmetric tensors take the same schematic form after the consideration of the electric/magnetic SSS ansatz:
\begin{equation}
\begin{split}
F_{\mu \nu}&=q(r) \tau^t_{[\mu} \rho^r_{\nu]}+p(r) \sigma^\theta_{[\mu} \sigma^\phi_{\nu]}\,,\\
\mathcal{M}_{\mu \nu}&=\tilde{q}(r) \tau^t_{[\mu} \rho^r_{\nu]}+\tilde{p}(r) \sigma^\theta_{[\mu} \sigma^\phi_{\nu]}\, ,
\end{split}
\end{equation}
where $q(r), \tilde{q}(r), p(r)$ and $\tilde{p}(r)$ are some radial functions and where we defined the projectors
\begin{equation}
\tau_{\mu}^\nu=\delta_\mu^t \delta_t^\nu\, , \quad \rho_\mu^\nu= \delta_\mu^r \delta_r^\nu\, , \quad \sigma_\mu^\nu=\sum_{i=1}^2 \delta_\mu^i \delta_i^\nu\, ,
\end{equation}
where the index $i$ runs over the angular coordinates.

On the other hand, the Riemann tensor, and thence the $\tensor{P}{_{\mu \nu\alpha \beta}}$ tensor, only have the following type of components\footnote{One may see, upon use of the conditions $L_{k_{(i)}} R_{\mu \nu \rho \sigma}=L_{k_{(i)}} P_{\mu \nu \rho \sigma}=0$ for $i={1,\dots,4}$, that any of the two expressions given at \eqref{eq:decomppr} represents the most general tensor with the same symmetries as the Riemann  and consistent with the static and spherical symmetry.} when evaluated on \req{eq:sss1} \cite{Deser:2005pc}:
\begin{equation}
\label{eq:decomppr}
\begin{split}
\tensor{R}{_{\mu \nu}^{\alpha \beta}}&= A(r) \tau_{[\mu}^{[\alpha} \rho_{\nu]}^{\beta]}+ B(r) \tau_{[\mu}^{[\alpha} \sigma_{\nu]}^{\beta]}+C(r)  \rho_{[\mu}^{[\alpha} \sigma_{\nu]}^{\beta]}+D(r) \sigma_{[\mu}^{[\alpha} \sigma_{\nu]}^{\beta]}\, ,\\
\tensor{P}{_{\mu \nu}^{\alpha \beta}}&= \tilde{A}(r) \tau_{[\mu}^{[\alpha} \rho_{\nu]}^{\beta]}+ \tilde{B}(r) \tau_{[\mu}^{[\alpha} \sigma_{\nu]}^{\beta]}+\tilde{C}(r)  \rho_{[\mu}^{[\alpha} \sigma_{\nu]}^{\beta]}+\tilde{D}(r) \sigma_{[\mu}^{[\alpha} \sigma_{\nu]}^{\beta]}\, ,
\end{split}
\end{equation}
where we used the projectors defined above and where $A(r), \tilde{A}(r), B(r) \dots$ are certain radial functions. Taking into account \eqref{eq:decomppr}, we check by direct computation that both $\tensor{P}{_{(\mu}^{\alpha \beta \gamma}} R_{\nu) \alpha \beta \gamma}$ and $\nabla^\sigma \nabla^\rho P_{(\mu| \sigma | \nu) \rho}$ have vanishing off-diagonal components.

Finally we must check that the $\theta \theta$ and $\phi \phi$ components of the Einstein equations are satisfied once the $tt$ and $rr$ are. For that, let us think of the Bianchi identity associated to the diffeomorphism invariance of any action of the form \eqref{eq:vmga}. Evidently this Bianchi identity is different from the usual $\nabla^\mu \left ( R_{\mu \nu}-\frac{1}{2} g_{\mu \nu} R \right )=0$ (which holds too) since there are higher-order terms in the curvature in addition to non-trivial couplings to electromagnetism. However we can equally apply the Second Noether Theorem to \eqref{eq:vmga} in order to obtain the following off-shell identity:
\begin{equation}
\nabla_\mu \left ( \frac{\delta I}{\delta g_{\mu \nu}} \right ) +H \left ( \frac{\delta I}{\delta A_\nu} \right )=0\,,
\end{equation}
where $H \left ( \frac{\delta I}{\delta A_\nu} \right )$ is a certain function which depends on the generalized Maxwell's equation of motion and vanishes when $\frac{\delta I}{\delta A_\nu}=0$. Therefore, if the vector field is a solution of the generalized Maxwell equation \eqref{eq:MaxwellEq} --- for instance, the magnetic vector given by \req{Fmagnetic} --- we obtain that 
%This identity can be obtained explicitly, after some algebra, by taking the divergence of the equation \eqref{eq:EinsteinEq}. Consequently, if the vector field is a solution of the generalized Maxwell equation \eqref{eq:MaxwellEq} --- for instance, the magnetic vector given by \req{Fmagnetic} --- we obtain that 
\begin{equation}
\left. \nabla_\mu \left ( \frac{\delta I}{\delta g_{\mu \nu}} \right )  \right \vert_{F=F_{\rm sol}}=0\,.
\end{equation}
Thus, we can assume the Bianchi identity $\nabla_{\mu}\mathcal{E}^{\mu\nu}=0$ once the Maxwell equation is solved. Evaluating this identity on the SSS metric \req{eq:sss1} and taking into account that $\mathcal{E}_{\mu\nu}$ has no off-diagonal components, we may find after some algebra that the $\nu=r$ component of the divergence of the Einstein equation \eqref{eq:EinsteinEq} takes the form:
\begin{equation}
	\frac{ d \mathcal{E}^{rr}}{dr}+\left(\frac{2}{r}-\frac{1}{2}f^{-1}f'+N^{-1} N'\right)\mathcal{E}^{rr}+\left (f^2 N N'+ \frac{1}{2} N^2 f f' \right )\mathcal{E}^{tt}-\frac{f}{r}g_{ij}\mathcal{E}^{ij}=0\, ,
\end{equation}
where $i,j$ are the angular components. Since due to spherical symmetric $\mathcal{E}^{\theta\theta}$ and $\mathcal{E}^{\phi\phi}$ are proportional to each other, then this identity implies that whenever $\mathcal{E}^{rr}=\mathcal{E}^{tt}=0$, then $\mathcal{E}^{\theta\theta}=\mathcal{E}^{\phi\phi}=0$, and the equations of motion are solved.

\section{All Electromagnetic (Generalized) Quasitopological Gravities of the form $RF^2$ and  $R^2F^2$}\label{app:a}

In this appendix we are going to build all Electromagnetic (generalized) quasi-topological gravities constructed by linear combinations of terms up to quadratic order both in the Riemann curvature tensor $R_{\mu \nu \rho \sigma}$ and in the gauge field strength $F_{\mu \nu}$. For that, we first classify all E(G)Qs $RF^2$, made up of scalar terms with one Riemann and two field strength and, afterwards, we proceed analogously for E(G)Q theories $R^2F^2$, whose constituent terms contain exactly two Riemann tensors and two field strengths.

\subsection{$RF^2$ theories}

Our first task is to find a set of  diffeomorphism-invariant terms which span all possible scalars of the form $RF^2$. This can be done straightforwardly and we find the following basis of invariants containing one Riemann tensor and two field strengths:
\begin{equation}
\mathcal{I}_1= R F^2\, , \quad \mathcal{I}_2= R_{\mu \nu} F^{\mu \alpha} \tensor{F}{^\nu_\alpha}\, , \quad \mathcal{I}_3= R_{\mu \nu \rho \sigma} F^{\mu \nu} F^{\rho \sigma}\,. 
\end{equation}
Now we build the Lagrangian density
\begin{equation}
\mathcal{L}=R+\ell^{2}\sum_{i=1}^{3} a_i \mathcal{I}_i\, , \quad a_i \in \mathbb{R}
\label{eq:applag}
\end{equation}
and wonder when the corresponding theory belongs to the Electromagnetic (Generalized) Quasitopological type. For that, we just need to check when the Definition \ref{def:EGQG} is fulfilled. Setting $L_f=r^2 \left. \mathcal{L}\right \vert_{d s^2_f, F^m}$, where $d s^2_f$ and $F^m$ are given by \eqref{eq:dsf}  \eqref{Fmagnetic} respectively, we have that
\begin{equation}
\frac{\partial L_f}{\partial f}-\frac{\mathrm{d}}{\mathrm{d} r} \frac{\partial L_f}{\partial f'}+\frac{\mathrm{d}}{\mathrm{d} r^2 } \frac{\partial L_f}{\partial f''}=-\frac{4 P^2 \ell^2 (10a_1+2a_2+a_3)}{r^4}\,.
\end{equation}
For the theory to be of the (Generalized) Quasitopological type, we must ensure that the latter expression vanishes. This is accomplished by
\begin{equation}
a_3=-10a_1-2a_2\,.
\end{equation}
Now the equation of motion for the metric function $f(r)$ is obtained by evaluating the Lagrangian \eqref{eq:applag} on the general SSS ansatz \eqref{eq:sss1} with a magnetic vector \eqref{Fmagnetic}, varying the subsequent action with respect to $N$ and, finally, imposing the condition $N=1$. Through this procedure, one finds the following equation of motion for $f(r)$:
\begin{equation}
\frac{\mathrm{d}}{\mathrm{d}r}\left [2r(1-f)+\frac{2 P^2 \ell^2 (6 a_1+a_2+2 a_1 f(r)}{r^3} \right ]=0\,.
\end{equation}
This equation can be directly integrated to yield
\begin{equation}
1-f-\frac{2M}{r}+\frac{ P^2 \ell^2}{r^4}(6 a_1+a_2+2 a_1 f(r))=0\,,
\label{eq:appeom12}
\end{equation}
where $M$ is an integration constant appropriately chosen to be identified with the mass, as done in the main text. 

There are two important conclusions to extract from Eq. \eqref{eq:appeom12}. Firstly, we recognize precisely the same structure as in Eq. \eqref{eq:eomf}, if we limit ourselves to the Einstein-Hilbert term and the terms with $n=1, m=2$. Secondly, we check that the set of EGQs and EQs coincide for theories of the form $RF^2$, since Eq. \eqref{eq:appeom12} is algebraic. This property does not hold generally of course and is very particular of $RF^2$ theories. As a matter of fact, the special properties of these Lagrangians had been previously noticed in the literature \cite{MuellerHoissen:1988bp,Balakin:2015gpq,Balakin:2016mnn}. 

 \subsection{$R^2F^2$ theories}
Again, first of all we shall concentrate on finding a set of invariants spanning all possibles scalars built out with precisely two Riemanns and two field strengths. After some work, it is possible to choose such set to be\footnote{Note that we are not claiming that all terms in this set are linearly independent.}:
\begin{equation}
\begin{split}
&\mathcal{I}_1= R^2 F^2\, , \quad \mathcal{I}_2=R R_{\mu \nu} F^{\mu \alpha} \tensor{F}{^\nu_\alpha} \, , \quad \mathcal{I}_3= R_{\mu \nu} R^{\mu \nu} F^2\, , \quad \mathcal{I}_4=R_{\mu \nu} \tensor{R}{_\alpha^\nu} F^{\mu \beta} \tensor{F}{^\alpha_\beta}\, , \\& \mathcal{I}_5=R_{\mu \nu} R_{\alpha \beta} F^{\mu \alpha} F^{\nu \beta}\,, \quad \mathcal{I}_6= R R_{\mu \nu \rho \sigma} F^{\mu \nu} F^{\rho \sigma} \, , \quad\mathcal{I}_7=R_{\mu \nu} \tensor{R}{^{\mu \alpha \nu \beta}}F_{\alpha \rho } \tensor{F}{_\beta^\rho}\, , \\& \mathcal{I}_8=R_{\mu \nu} \tensor{R}{^\mu_{\alpha \beta \sigma}} F^{\beta \sigma} \tensor{F}{^{\alpha \nu}} \, , \quad \mathcal{I}_9=R_{\mu \nu} \tensor{R}{^\mu_{\alpha \beta \sigma}} F^{\alpha \beta} \tensor{F}{^{\sigma \nu}}\, , \quad \mathcal{I}_{10}=R_{\mu \nu \rho \sigma} R^{\mu \nu \rho \sigma} F^2\, \\& \mathcal{I}_{11}= R_{\mu \nu \rho \alpha} R^{\mu \nu \rho \beta}  F^{\alpha \lambda} F_{\beta \lambda} \, , \quad \mathcal{I}_{12}=R_{\mu \nu \rho \sigma} R^{\mu \nu \alpha \beta} F^{\rho \sigma} F_{\alpha \beta} \, , \quad \mathcal{I}_{13}= R_{\mu \nu \rho \sigma} R^{\mu \nu \alpha \beta} \tensor{F}{^\rho_\alpha} \tensor{F}{^\sigma_\beta} \, , \\& \mathcal{I}_{14}=R_{\mu \nu \rho \sigma} R^{\mu \alpha \rho \beta} F^{\nu \sigma} F_{\alpha \beta} \, , \quad \mathcal{I}_{15}=R_{\mu \nu \rho \sigma} R^{\mu \alpha \rho \beta} \tensor{F}{^\nu_\alpha} \tensor{F}{^\sigma_\beta}\, .
\end{split}
\end{equation}
Proceeding in the same way as with $RF^2$ theories, now we consider the Lagrangian density
\begin{equation}
\mathcal{L}=R+\ell^{4}\sum_{i=1}^{15} b_i \mathcal{I}_i\, , \quad b_i \in \mathbb{R}
\label{eq:applag}
\end{equation}
and investigate when the corresponding theory belongs to the E(G)Q type. Defining as before $L_f=r^2 \left. \mathcal{L}\right \vert_{d s^2_f, F^m}$, we have that
\begin{equation}
\frac{\partial L_f}{\partial f}-\frac{\mathrm{d}}{\mathrm{d} r} \frac{\partial L_f}{\partial f'}+\frac{\mathrm{d}}{\mathrm{d} r^2 } \frac{\partial L_f}{\partial f''}= \frac{P^2 \ell^4 }{r^6} \left (\mathcal{A}_1-\mathcal{A}_1  f+\mathcal{A}_2 r f'+\mathcal{A}_3 r^2 f''-4\mathcal{A}_4 r^3 f^{(3)} +\mathcal{A}_4 r^4 f^{(4)}\right)\,,
\end{equation}
where we have defined
\begin{equation}
\begin{split}
\mathcal{A}_1&=-2(168 b_1+8b_{10}+4b_{11}+8b_{12}+4b_{13}+2b_{14}+54 b_2+24 b_3+12 b_4\\&+12 b_5+88 b_6+7(b_7-2 b_8+b_9))\,, \\
\mathcal{A}_2&=4(96 b_1+2(8b_{10}+2b_{11}+b_{15}+9 b_2+2(7b_3+b_4+b_5-2b_6))+7b_7)\,, \\
\mathcal{A}_3&=-2(36b_1-4b_{10} +2b_{11}+b_{15}+9b_2+2(4b_3+b_4+b_5-2b_6))-7b_7\, , \\
\mathcal{A}_4&=4(b_1+b_{10})+2b_3\, , \\
\end{split}
\end{equation}
The theory is a (Generalized) Quasitopological one if all $\mathcal{A}_i$ vanish simultaneously. Such a system of linear equations is solved by:
\begin{equation}
\begin{split}
b_3=&-2b_1-2b_{10} \, , \quad 7 b_7=-40 b_1+40 b_{10}-4b_{11}-2 b_{15}-18 b_2-4b_4-4b_5+8b_6\, , \\7 b_9&=-80b_1-8b_{12}-4b_{13}-2 b_{14}+2 b_{15}-36 b_2 -8 b_4-8 b_5-96 b_6+14 b_8\,. 
\end{split} 
\end{equation}
Imposing these constraints, one obtains the generic expression for any EGQ constructed out of terms with two Riemanns and two field strengths. However, we still need to figure out which of these EGQs are actually quasitopological.

For that, we must learn when the equation for $f(r)$ is algebraic. As aforementioned, this equation is derived after evaluating the Lagrangian \eqref{eq:applag} on our magnetic SSS ansatz, varying the subsequent action respect to $N$ and afterwards setting $N=1$. One obtains:

% The equation of motion for the metric function $f(r)$ is obtained by evaluating the Lagrangian \eqref{eq:applag} on the general SSS ansatz \eqref{eq:sss1} with a magnetic vector \eqref{Fmagnetic}, varying the subsequent action with respect to $N$ and, finally, imposing the condition $N=1$. Through this procedure, one finds the following equation of motion for $f(r)$:
\begin{equation}
\frac{\mathrm{d}}{\mathrm{d}r}\left [2r(1-f)+ \frac{P^2 \ell^4}{7 r^5} \left (\mathcal{B}_1+\mathcal{B}_2 f+\mathcal{B}_3 f^2+6\mathcal{B}_4 rf f'+ \mathcal{B}_4 r^2 f'^2-2\mathcal{B}_4 r^2f'' \right ) \right ]=0\, ,
\end{equation}
where 
\begin{equation}
\begin{split}
\mathcal{B}_1&=2(24 b_1-8 b_{10}-2 b_{11}-4 b_{12}-2 b_{13}-b_{14}+8b_2+b_4+b_5+12b_6)\, , \\
\mathcal{B}_2&=56(4b_1+b_2+2b_6)\, , \\
\mathcal{B}_3&=-2(136 b_1-8 b_{10}-2b_{11}-4 b_{12}-2 b_{13}-b_{14}+36 b_2+b_4+b_5+68b_6\, , \\
\mathcal{B}_4&=-8 b_1+8 b_{10}+2 b_{11}+b_{15}+2 b_2+2 b_4 +2 b_5-4 b_6\,.
\end{split} 
\end{equation}
EQs are characterized by having an algebraic equation of motion for the metric function $f(r)$. Interestingly enough, this is achieved if we just impose the vanishing of $\mathcal{B}_4$. Therefore, setting $\mathcal{B}_4=0$, we get the most general form for the equation of motion of $f(r)$ in any EQ built out of linear combinations of scalars with at most two Riemann curvature tensors and two gauge field strengths. This equation reads as follows:
\begin{equation}
\frac{\mathrm{d}}{\mathrm{d}r}\left [2r(1-f)+ \frac{
2P^2 \ell^4 }{7 r^5} (1-f)(\mathcal{C}_1+\mathcal{C}_2 f) \right ]=0\,, 
\label{eq:appefr2f2}
\end{equation}
where
\begin{equation}
\begin{split}
\mathcal{C}_1&=16 b_{10}+4b_{11}-4b_{12}-2 b_{13}-b_{14}+3 b_{15}+14 b_2+7 b_4+7 b_5\, , \\
\mathcal{C}_2&=128 b_{10}+32 b_{11}-4 b_{12}-2b_{13}-b_{14}+17 b_{15} +70 b_2 +35 b_4+35 b_5\, .
\end{split}
\end{equation}
Upon direct integration of the previous expression, choosing appropriately the constant of integration $M$, we end up with 
\begin{equation}
1-f-\frac{2M}{r}+ \frac{
P^2 \ell^4 }{7 r^6} (1-f)(\mathcal{C}_1+\mathcal{C}_2 f) =0\, ,
\end{equation}
and we recognize the same structure as in Eq. \eqref{eq:eomf} after restricting ourselves to those terms with $n=2, m=1$. Hence we have proven that the equation for $f(r)$ of the most general EQ constructed from terms with at most two Riemann tensors and two gauge field strengths is indeed represented by Eq. \eqref{eq:eomf}, after an appropriate choice of couplings.

\bibliography{Gravities}
\bibliographystyle{JHEP-2}
\label{biblio}

\end{document}